\documentclass[a4paper,fleqn]{cas-dc}

\usepackage[authoryear,longnamesfirst]{natbib}

\def\tsc#1{\csdef{#1}{\textsc{\lowercase{#1}}\xspace}}
\tsc{WGM}
\tsc{QE}
\usepackage{xcolor}
\usepackage{enumitem}
\usepackage{amsmath,amssymb}
\usepackage{placeins}
\usepackage[normalem]{ulem}
\usepackage{tikz}
\usetikzlibrary{arrows.meta,positioning,shapes.geometric,fit,calc,decorations.pathreplacing}

\newcommand{\cpssd}{CP-SSD}
\newcommand{\kbkssd}{KB-KSSD}
\newcommand{\jagt}{JAGT}

\begin{document}
\let\WriteBookmarks\relax
\def\floatpagepagefraction{1}
\def\textpagefraction{.001}

\shorttitle{Automatic SSD Recommendation}    

\shortauthors{Beseda et al.}  

\title [mode = title]{Towards Automated Recommendation of Steady-State Detection Algorithms for Software Performance Benchmarks}  


%\tnotetext[1]{Working manuscript draft.} 


\author{Martin Beseda}[orcid=0000-0001-5792-2872]

\cormark[1]


\ead{martin.beseda@univaq.it}


\credit{Conceptualization, Methodology, Software, Investigation, Writing}

\affiliation{organization={University of L'Aquila},
            addressline={via Vetoio (Coppito), 1}, 
            city={L'Aquila},
            postcode={67100},
            country={Italy}}

\author{Vittorio Cortellessa}[orcid=0000-0002-4507-464X]

% Footnote of the second author

% Email id of the second author
\ead{vittorio.cortellessa@univaq.it}

\author{Federico Di Menna}[orcid=0009-0003-5834-6389]

% Footnote of the second author

% Email id of the second author
\ead{federico.dimenna@graduate.univaq.it}


\author{Daniele Di Pompeo}[orcid=0000-0003-2041-7375]

% Footnote of the second author

% Email id of the second author
\ead{daniele.dipompeo@univaq.it}


\author{Luca Traini}[orcid=0000-0003-3676-0645]

% Footnote of the second author

% Email id of the second author
\ead{luca.traini@univaq.it}


\credit{Supervision, Methodology, Writing -- review and editing}

% \affiliation[2]{organization={TODO Institution},
%             addressline={TODO Address}, 
%             city={TODO City},
% %          citysep={}, % Uncomment if no comma needed between city and postcode
%             country={TODO Country}}

%\cortext[1]{Corresponding author}



\begin{abstract}
Steady-state detection is a central step in software performance benchmarking,
yet different detection methods rely on different assumptions and may perform
best on different execution profiles. This paper formulates steady-state
detection as an algorithm-recommendation problem: given a benchmark time
series, predict which of two complementary detectors is more appropriate.
The first, CP-SSD, identifies steady state through changepoint-based
segmentation of the execution trace, whereas the second, KB-KSSD, uses a
kernel-based approach to identify stable behaviour. We propose a data-driven recommender that characterises benchmark time series
using automatically extracted features and combines supervised prediction with
information derived from groups of time series exhibiting similar behaviour.
The approach is evaluated on 407 labelled benchmark time series for which one
of the two detectors can be identified as preferable with respect to a
reference steady-state index. The proposed recommender selects the better
detector correctly for 72.97\% of the series, compared with 62.65\% when
always selecting the detector that performs best most often, corresponding to
an absolute improvement of 10.32 percentage points. Paired statistical testing
and bootstrap confidence intervals support this improvement. 
A comparison with a recommender using time-series features alone shows that
the cluster-derived information mainly improves interpretability rather than
predictive accuracy. Recommendations also remain highly stable when the
unsupervised representation is learned from either the labelled subset or the
full collection of 5,860 time series. These results demonstrate the feasibility
of per-series steady-state detector recommendation while highlighting the need
for broader labelled data, particularly for execution profiles favouring the
changepoint-based detector.
\end{abstract}


\begin{highlights}
\item We formulate steady-state detection as an automatic algorithm-recommendation task.
\item On 407 steady benchmark time series with judgment-aggregated ground truth, our TSFresh-plus-cluster-signature recommender achieves 72.97\% accuracy, compared with 62.65\% when always selecting Kernel-based Kelly's Steady-State Detection.
\item Exact McNemar testing and bootstrap confidence intervals show that the gain over always selecting the majority-winning detector is statistically significant and robust to project-level resampling.
\item Recommendations are highly stable when clustering is learned either from the 407 labelled steady series or from all 5,860 available time series, with 98.03\% agreement.
\end{highlights}

%\nocite{*}

% Keywords

\maketitle

% Main text
\section{Introduction}\label{sec:introduction}
Reliable performance benchmarking requires distinguishing measurements affected
by unstable execution behaviour from those that are sufficiently stable to be
used for performance inference. This distinction is particularly important for
managed runtimes, where just-in-time compilation, garbage collection, resource allocation, and other runtime mechanisms can cause execution times to change substantially during the initial part of a
benchmark run \citep{georges2007,kalibera2013}. In many executions,
this behaviour can be interpreted as a transition from an initial
\emph{warm-up} region towards a more stable \emph{steady-state} region.
However, such a transition is not necessarily sharp or even guaranteed to
occur: benchmark traces may contain temporary plateaus, repeated transitions,
outliers, or long-term non-stationarity \citep{barrett2017,traini2023}.

Identifying the beginning of a steady state is therefore an important and
non-trivial part of performance analysis. If the transition is detected too
early, unstable observations may be included in the final performance estimate
and bias the reported result. Conversely, an overly conservative detector may
discard useful observations, unnecessarily extend benchmark execution, or
reduce the amount of steady-state data available for subsequent analysis.
Consequently, steady-state detection (SSD) has received increasing attention in
software performance benchmarking, where several approaches have been proposed, ranging from
changepoint-based segmentation to statistical and probabilistic tests
\citep{beseda2026,barrett2017,traini2023,bethea2019applied,boesler2017steady,cao1996statistically,wu2016online}.

An issue that comparatively receives less attention is that SSD methods
embody different assumptions about what constitutes the transition to steady
behaviour. Changepoint-based approaches identify structural changes in the
observed execution trace and infer a final stable region from the resulting
segmentation. In contrast, other methods may evaluate local smoothness,
probabilistic steadiness, or specific transition patterns. These differences are
not merely methodological: they can lead different detectors to be preferable on different execution profiles.

This is particularly visible when comparing two recent SSD approaches used in
this study. The first is a changepoint-based approach introduced by Traini
et al., which we denote as \cpssd{} \citep{traini2023}. The second is
\kbkssd{}, introduced in our previous work \citep{beseda2026}, which combines robust preprocessing,
kernel-based transition detection, and a probabilistic steady-state criterion
adapted from process monitoring. 

Indeed, on the benchmark
collection considered in that study, \kbkssd{} achieved lower total
detection error with respect to judgment-aggregated ground truth (JAGT), which
provides the human-selected reference steady-state indices against which automatic detections are evaluated, and fewer false-negative detections than \cpssd{}.
Nevertheless, the total result does not imply that \kbkssd{} is the better
choice for every individual execution. In the 407 steady JAGT time series used in
the present work, \kbkssd{} provides the better steady-state estimate for 255
of them, whereas \cpssd{} works better for the remaining 152. Thus, always applying
a globally stronger detector is not always the best choice, as it can be applied to a substantial number of benchmark
executions where another method could produce more accurate results.

This observation motivates a different formulation of the SSD problem.
Instead of asking which detector is best \emph{on average}, we ask whether the
characteristics of an individual benchmark time series can be used to predict
which detector would work better on that particular execution. The problem is
therefore naturally related to \emph{per-instance algorithm selection}, where
features describing an instance are used to choose the algorithm expected to
perform best on that instance \citep{rice1976,kerschke2019}. In
the present context: an ``instance'' is a benchmark execution trace, the
algorithm portfolio contains \cpssd{} and \kbkssd{}, and the target is the
detector producing the smaller absolute error relative to the reference JAGT
steady-state indices. More broadly, this formulation represents an application of artificial
intelligence to software performance analysis, using machine learning to adapt
the choice of SSD method to the characteristics of each execution trace.

The central challenge is therefore to construct a representation of an
execution trace that contains information about the relative performance of the
two SSD methods. We begin by extracting a broad set of time-series descriptors
with TSFresh \citep{christ2018}, covering distributional, entropy-based,
frequency-domain, temporal, and other characteristics. These descriptors are
used in two complementary branches of the recommendation pipeline.
The first branch searches for structure among execution traces. We reduce the
TSFresh representation using Principal Component Analysis (PCA)
\citep{abdi2010principal} and Uniform Manifold Approximation and Projection
(UMAP) \citep{mcinnes2018}, and subsequently identify groups of similar traces
using HDBSCAN \citep{mcinnes2017hdbscan}. Clusters containing labelled series
are then characterised by whether their members tend to favour \cpssd{} or
\kbkssd{}. This information is converted into cluster-signature similarity
features that describe how closely a new execution resembles regions associated
with either detector.

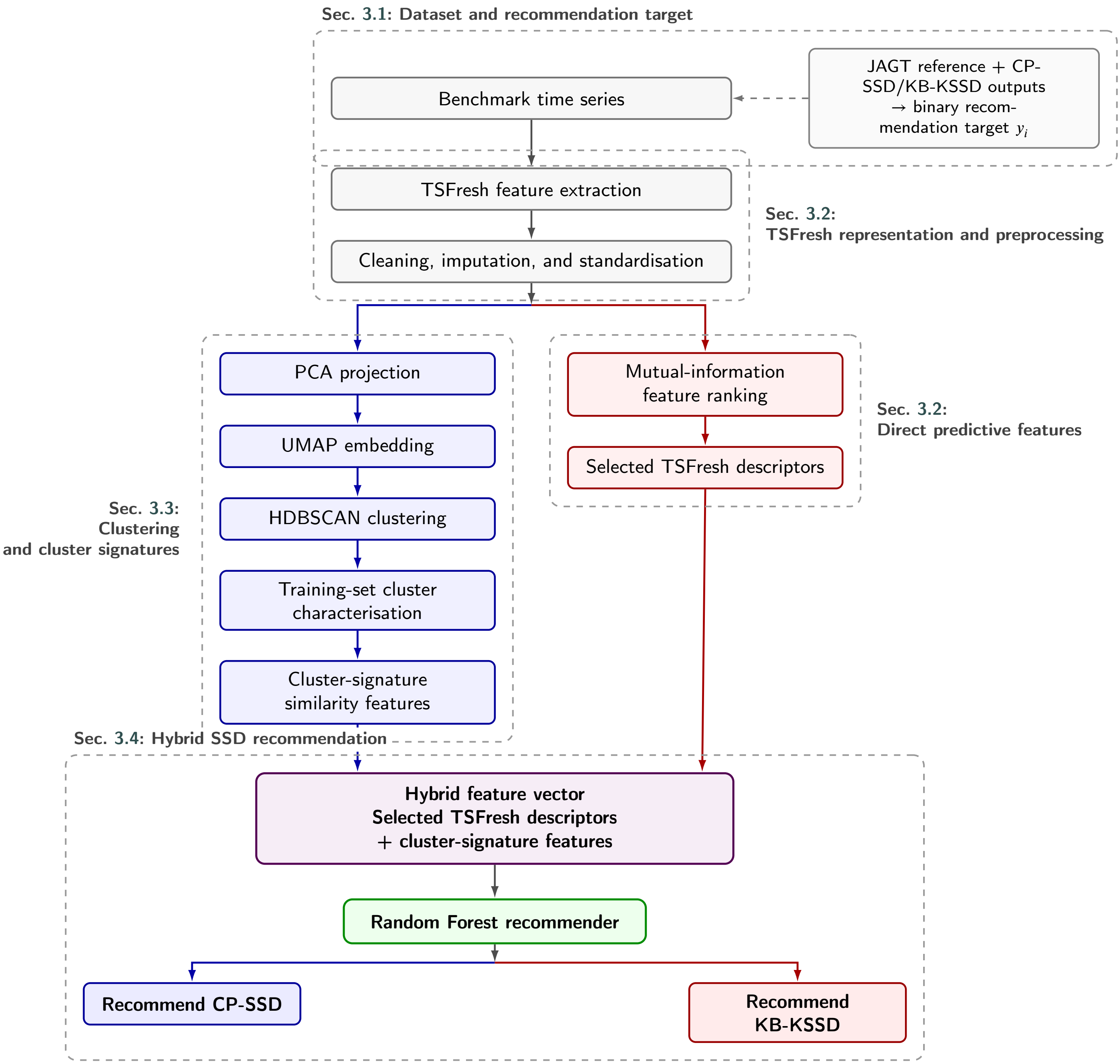
\begin{figure*}[bt]
\centering
\begin{tikzpicture}[
    node distance=0.48cm and 1.35cm,
    every node/.style={font=\small},
    flow/.style={
        -{Latex[length=2mm,width=1.3mm]},
        line width=0.85pt,
        draw=black!70
    },
    auxflow/.style={
        -{Latex[length=1.7mm,width=1.1mm]},
        line width=0.7pt,
        draw=black!55,
        dashed
    },
    structureflow/.style={
        -{Latex[length=2mm,width=1.3mm]},
        line width=0.9pt,
        draw=blue!65!black
    },
    predictiveflow/.style={
        -{Latex[length=2mm,width=1.3mm]},
        line width=0.9pt,
        draw=red!65!black
    },
    shared/.style={
        rectangle,
        rounded corners=3pt,
        draw=black!55,
        fill=black!3,
        line width=0.8pt,
        text width=6.2cm,
        minimum height=0.68cm,
        align=center,
        inner sep=5pt
    },
    target/.style={
        rectangle,
        rounded corners=3pt,
        draw=black!50,
        fill=black!2,
        line width=0.75pt,
        text width=4.4cm,
        minimum height=0.68cm,
        align=center,
        inner sep=5pt,
        font=\footnotesize
    },
    structure/.style={
        rectangle,
        rounded corners=3pt,
        draw=blue!60!black,
        fill=blue!6,
        line width=0.85pt,
        text width=4.15cm,
        minimum height=0.68cm,
        align=center,
        inner sep=5pt
    },
    predictive/.style={
        rectangle,
        rounded corners=3pt,
        draw=red!60!black,
        fill=red!6,
        line width=0.85pt,
        text width=4.15cm,
        minimum height=0.68cm,
        align=center,
        inner sep=5pt
    },
    merge/.style={
        rectangle,
        rounded corners=4pt,
        draw=violet!65!black,
        fill=violet!7,
        line width=0.95pt,
        text width=7.4cm,
        minimum height=0.82cm,
        align=center,
        inner sep=6pt,
        font=\small\bfseries
    },
    classifier/.style={
        rectangle,
        rounded corners=4pt,
        draw=green!55!black,
        fill=green!7,
        line width=0.95pt,
        text width=4.6cm,
        minimum height=0.72cm,
        align=center,
        inner sep=5pt,
        font=\small\bfseries
    },
    outputcp/.style={
        rectangle,
        rounded corners=3pt,
        draw=blue!60!black,
        fill=blue!8,
        line width=0.85pt,
        text width=3.2cm,
        minimum height=0.68cm,
        align=center,
        inner sep=5pt,
        font=\small\bfseries
    },
    outputkb/.style={
        rectangle,
        rounded corners=3pt,
        draw=red!60!black,
        fill=red!8,
        line width=0.85pt,
        text width=3.2cm,
        minimum height=0.68cm,
        align=center,
        inner sep=5pt,
        font=\small\bfseries
    },
    lanehead/.style={
        font=\small\bfseries,
        align=center
    },
    sectionbox/.style={
        rectangle,
        rounded corners=5pt,
        draw=black!40,
        dashed,
        line width=0.75pt,
        inner sep=8pt
    },
    sectionlabel/.style={
        font=\footnotesize\bfseries,
        fill=white,
        text=black!75,
        inner sep=2pt
    }
]

% ------------------------------------------------------------------
% Section 3.1: benchmark input and recommendation target
% ------------------------------------------------------------------

\node[shared] (input) {Benchmark time series};

\node[target, right=1.25cm of input] (target)
{JAGT reference + CP-SSD/KB-KSSD outputs\\
$\rightarrow$ binary recommendation target $y_i$};

\draw[auxflow] (target.west) -- (input.east);

% ------------------------------------------------------------------
% Section 3.2: shared TSFresh representation and preprocessing
% ------------------------------------------------------------------

\node[shared, below=0.78cm of input] (tsfresh)
{TSFresh feature extraction};

\node[shared, below=of tsfresh] (preprocess)
{Cleaning, imputation, and standardisation};

\draw[flow] (input) -- (tsfresh);
\draw[flow] (tsfresh) -- (preprocess);

% ------------------------------------------------------------------
% Two feature-construction branches
% ------------------------------------------------------------------

\coordinate (branchorigin) at ([yshift=-0.95cm]preprocess.south);

\node[
    lanehead,
    text=blue!65!black,
    anchor=south
] (structurehead)
at ([xshift=-2.85cm]branchorigin)
{};%{Structure-discovery branch};

\node[
    lanehead,
    text=red!65!black,
    anchor=south
] (predictivehead)
at ([xshift=2.85cm]branchorigin)
{};%{Predictive-feature branch};

% ------------------------------------------------------------------
% Section 3.3: structure-discovery branch
% ------------------------------------------------------------------

\node[structure, below=0.18cm of structurehead] (pca)
{PCA projection};

\node[structure, below=of pca] (umap)
{UMAP embedding};

\node[structure, below=of umap] (hdbscan)
{HDBSCAN clustering};

\node[structure, below=of hdbscan] (characterize)
{Training-set cluster characterisation};

\node[structure, below=of characterize] (signature)
{Cluster-signature similarity features};

\draw[structureflow] (pca) -- (umap);
\draw[structureflow] (umap) -- (hdbscan);
\draw[structureflow] (hdbscan) -- (characterize);
\draw[structureflow] (characterize) -- (signature);

% ------------------------------------------------------------------
% Section 3.2: direct predictive-feature branch
% ------------------------------------------------------------------

\node[predictive, below=0.18cm of predictivehead] (mi)
{Mutual-information feature ranking};

\node[predictive, below=of mi] (selected)
{Selected TSFresh descriptors};

\draw[predictiveflow] (mi) -- (selected);

% ------------------------------------------------------------------
% Split the shared representation into the two branches
% ------------------------------------------------------------------

\coordinate (split) at ([yshift=-0.36cm]preprocess.south);

\draw[flow] (preprocess.south) -- (split);
\draw[structureflow] (split) -| (pca.north);
\draw[predictiveflow] (split) -| (mi.north);

% ------------------------------------------------------------------
% Section 3.4: hybrid representation and final recommendation
% ------------------------------------------------------------------

\node[
    merge,
    below=0.78cm of signature,
    xshift=2.25cm
] (hybrid)
{Hybrid feature vector\\
Selected TSFresh descriptors + cluster-signature features};

\coordinate (mergeleft)
at ([xshift=-2.25cm]hybrid.north);

\coordinate (mergeright)
at ([xshift=3.4cm]hybrid.north);

\draw[structureflow] (signature.south) -- (mergeleft);
\draw[predictiveflow] (selected.south) -- (mergeright);

\node[classifier, below=0.55cm of hybrid] (rf)
{Random Forest recommender};

\draw[flow] (hybrid) -- (rf);

\node[
    outputcp,
    below left=0.62cm and 0.68cm of rf
] (cp)
{Recommend CP-SSD};

\node[
    outputkb,
    below right=0.62cm and 0.68cm of rf
] (kb)
{Recommend KB-KSSD};

\coordinate (outsplit)
at ([yshift=-0.30cm]rf.south);

\draw[flow] (rf.south) -- (outsplit);
\draw[structureflow] (outsplit) -| (cp.north);
\draw[predictiveflow] (outsplit) -| (kb.north);

% ==================================================================
% Section-to-workflow mapping
% ==================================================================

% Section 3.1
\node[
    sectionbox,
    fit=(input)(target)
] (secDataset) {};

\node[
    sectionlabel,
    anchor=south west
] at ([xshift=2pt,yshift=1pt]secDataset.north west)
{Sec.~\ref{sec:dataset}: Dataset and recommendation target};

% Section 3.2 -- shared preprocessing
\node[
    sectionbox,
    fit=(tsfresh)(preprocess)
] (secPreprocessing) {};

% \node[
%     sectionlabel,
%     anchor=south west
% ] at ([xshift=2pt,yshift=1pt]secPreprocessing.north west)
% {Sec.~\ref{sec:feature-extraction}: TSFresh representation and preprocessing};

\node[
    sectionlabel,
    anchor=west,
    align=left
] at ([xshift=5pt]secPreprocessing.east)
{Sec.~\ref{sec:feature-extraction}:\\
TSFresh representation and preprocessing};

% Section 3.3 -- clustering branch
\node[
    sectionbox,
    fit=(pca)(umap)(hdbscan)(characterize)(signature)
] (secClustering) {};

% \node[
%     sectionlabel,
%     anchor=south west
% ] at ([xshift=2pt,yshift=1pt]secClustering.north west)
% {Sec.~\ref{sec:clustering}: Clustering and cluster signatures};

\node[
    sectionlabel,
    anchor=east,
    align=right
] at ([xshift=-8pt,yshift=4pt]secClustering.west)
{Sec.~\ref{sec:clustering}:\\
Clustering\\and cluster signatures};

% Section 3.2 -- direct predictive branch
\node[
    sectionbox,
    fit=(mi)(selected)
] (secDirect) {};

% \node[
%     sectionlabel,
%     anchor=south west
% ] at ([xshift=2pt,yshift=1pt]secDirect.north west)
% {Sec.~\ref{sec:feature-extraction}: Direct predictive features};

\node[
    sectionlabel,
    anchor=west,
    align=left
] at ([xshift=5pt]secDirect.east)
{Sec.~\ref{sec:feature-extraction}:\\
Direct predictive features};

% Section 3.4 -- hybrid recommendation
\node[
    sectionbox,
    fit=(hybrid)(rf)(cp)(kb)
] (secRecommendation) {};

\node[
    sectionlabel,
    anchor=south west
] at ([xshift=2pt,yshift=1pt]secRecommendation.north west)
{Sec.~\ref{sec:recommendation}: Hybrid SSD recommendation};

\end{tikzpicture}

\caption{Workflow of the hybrid SSD recommender and its correspondence to
the methodological stages described in Section~\ref{sec:methodology}. Section~\ref{sec:dataset} defines the benchmark
input and binary recommendation target. Section~\ref{sec:feature-extraction}
defines the shared TSFresh representation and preprocessing as well as the
direct predictive-feature branch. Section~\ref{sec:clustering} describes the
PCA--UMAP--HDBSCAN structure-discovery branch and the construction of
cluster-signature features. Section~\ref{sec:recommendation} combines the
direct and cluster-derived representations in the hybrid Random Forest
recommender. The dashed containers indicate these subsection boundaries.
Label-dependent operations, feature selection, and model fitting are performed
within the training portion of each evaluation split.}
\label{fig:workflow}
\end{figure*}

The second branch uses TSFresh descriptors directly as supervised predictive
features. We combine the selected direct descriptors with the cluster-signature
features and train a Random Forest classifier \citep{kulkarni2013random} to
recommend one of the two SSD methods. We refer to this as the \emph{hybrid}
recommender because its feature representation combines directly predictive
time-series descriptors with meta-features derived from the unsupervised
cluster structure. Figure~\ref{fig:workflow} summarises this pipeline.

This design lets us study two related questions. First, we ask whether
per-series recommendation can improve upon simple reference strategies,
including random selection and always selecting \kbkssd{}, which is the method
that wins most often in the labelled dataset. Second, we examine whether the
cluster-derived representation contributes useful information beyond the
directly predictive TSFresh descriptors. We therefore evaluate both the hybrid
recommender and an ablation that uses only the selected TSFresh features. The
purpose of the clustering component is not to assume that clustering must
improve predictive accuracy, but to determine whether method-specialised
regions exist and whether they provide an interpretable description of when the
two SSD approaches tend to succeed.

% This hybrid design serves two purposes. First, it allows us to test the
% practical question of whether SSD recommendation can outperform more natural
% strategies such as random selection or choosing the  more successful method every time.
% Second, it allows us to investigate whether unsupervised structure in the
% time-series representation provides information beyond the directly predictive
% TSFresh descriptors. We therefore evaluate both a direct TSFresh classifier and
% a hybrid model containing the additional cluster-signature features. This
% comparison is important because the objective of the present work is not to
% argue that clustering must necessarily improve classification accuracy, but to
% determine whether method-specialised regions exist and whether they can
% contribute to a humanly-interpretable recommendation framework.

% For every timeseries of the 407 steady ones, the recommendation target identifies
% whether \cpssd{} or \kbkssd{} produces the smaller error with respect to the
% reference steady-state index, including cases in which only one detector
% successfully identifies a steady state.

Based on these objectives, we investigate the following research questions:

\begin{description}[leftmargin=!,labelwidth=0.9cm]

\item[RQ1] \textbf{How often does each SSD method provide the better estimate,
and does method preference vary systematically across clusters of similar time
series?}

This question establishes whether the recommendation problem is meaningful in
the first place. If one detector dominates almost every series, per-instance
selection would offer little practical value. We therefore analyse both the
global distribution of \cpssd{} and \kbkssd{} wins and their distribution
within the clusters.

\item[RQ2] \textbf{Can the better SSD method be recommended automatically, and
does recommendation outperform random choice and/or majority-winning detector?}

We evaluate the predictive performance of our recommender against
equiprobable random selection, prevalence-matched random selection, and the
strong majority strategy of always selecting \kbkssd{}. We additionally compare
the hybrid model with a direct TSFresh classifier to isolate the incremental
contribution of the cluster-signature representation.

\item[RQ3] \textbf{What kinds of recommendation errors remain, particularly for
the minority CP-SSD class?}

Overall accuracy alone can obscure asymmetric behaviour between the two
classes. We therefore analyse the confusion matrix, balanced accuracy,
precision, recall, and F1-score, with particular attention to the ability to
recognise executions for which \cpssd{} is preferable.

\item[RQ4] \textbf{How sensitive are the recommendations to the population used
for unsupervised clustering?}

The primary clustering analysis is learned from the 407 labelled JAGT series,
whereas a secondary analysis derives the clustering structure from all 5,860
cached time series. We compare the resulting recommendations to determine
whether the downstream classifier depends strongly on the population used for
unsupervised structure discovery.

\end{description}

The main contributions of this paper are therefore:

\begin{itemize}[leftmargin=*]

\item We formulate SSD-method choice as a \emph{per-instance algorithm
recommendation problem}, rather than treating one detector as universally
preferable for all benchmark executions.

\item We construct a recommendation dataset from 407 decisive JAGT-labelled
time series, containing 152 \cpssd{} winners and 255 \kbkssd{} winners, using
explicit rules for detector failures and absolute-error comparisons.

\item We introduce a hybrid recommendation pipeline combining TSFresh
descriptors with similarity features derived from clusters of
behaviourally similar time series, providing both predictive and interpretable
information about detector preference.

\item We provide a repeated-holdout evaluation and a complementary
unique-series inferential analysis using exact paired hypothesis testing,
effect-size estimation, and bootstrap uncertainty quantification.

\item We show that automatic SSD recommendation improves substantially over
random and majority strategies, while also demonstrating that the present
cluster-signature representation does not produce a statistically supported
accuracy improvement over direct TSFresh features.

\item We show that the recommendations are highly robust to the population used
for clustering: clustering learned from the labelled JAGT subset and clustering
learned from all 5,860 cached time series produce 98.03\% identical final
recommendations.

\end{itemize}

The data, code, archived intermediate results, and replication
instructions accompanying this study are described in Section~\ref{sec:data}.

\section{Background and Related Work}\label{sec:related-work}
This section provides the background necessary to position the proposed SSD
recommendation framework. We first discuss steady-state behaviour and its
importance for reliable performance benchmarking, followed by the automatic
SSD methods that constitute the recommendation portfolio considered in this
work. We then relate the proposed formulation to the broader problem of
per-instance algorithm selection and finally review the time-series
representation, dimensionality-reduction, clustering, and classification
techniques used to construct the recommender.

\subsection{Managed-runtime benchmarking and steady-state behaviour}

Reliable microbenchmarking requires more than repeated execution: measurements
must be collected under a protocol that controls sources of variability and
separates transient behaviour from the region used for performance inference.
Georges et al. established a statistically rigorous methodology for Java
performance evaluation, while Kalibera and Jones subsequently showed how
careful experimental design can provide rigorous conclusions without
prohibitively long experiments \citep{georges2007,kalibera2013}. These concerns
are particularly important for managed runtimes, where just-in-time compilation,
garbage collection, adaptive optimisation, and other runtime mechanisms can
produce non-stationary execution profiles.

The conventional assumption that a virtual machine warms up and then remains at
a stable peak-performance state is not universally valid. Barrett et al. used an
automated changepoint-based analysis and showed that many benchmark--VM pairs do
not consistently exhibit the expected warm-up-to-steady-state pattern
\citep{barrett2017}. Traini et al. later investigated steady-state assessment in
Java microbenchmarking at a larger scale and showed that benchmarks
may fail to reach a steady state and that developer-specified warm-up durations
can be inaccurate \citep{traini2023}.

\subsection{Automatic steady-state detection}

The first candidate considered in this study is the change\-point-based procedure
used by Traini et al.; throughout this paper we refer to this procedure as
\cpssd{}. It builds on the changepoint formulation of Barrett et al. and uses
multiple-change detection to partition an execution trace into statistically
homogeneous regions \citep{barrett2017,traini2023}. The underlying PELT
algorithm finds an exact optimum for a penalised multiple-changepoint objective
while achieving linear computational cost under suitable conditions
\citep{killick2012}. Post-processing then removes negligible changes and the
final segments are inspected to determine whether a sufficiently stable region
has been reached.

The second candidate, \kbkssd{}, follows a different design. It combines robust
preprocessing, kernel-based transition detection, and a probabilistic
steady-state test adapted from Kelly and Hedengren's process-monitoring
formulation \citep{kelly2013}. The complete method was introduced for
performance time series by Beseda et al. and evaluated against CP-SSD and other
reference methods on a JAGT \citep{beseda2026}. That study reported a 14.5\% reduction in total absolute
steady-state-index error relative to CP-SSD and a lower false-negative rate.
Importantly for the present work, however, collection-level superiority does not
imply that one method is best for every benchmark execution. The JAGT-labelled
series contain substantial numbers of instances won by each method, which makes
per-instance selection meaningful.

The studies above evaluate SSD procedures as competing methods over benchmark
collections. The present work asks a different question: given the observed
characteristics of one execution trace, can the method that will produce the
smaller steady-state-index error be selected automatically? This shifts the
unit of decision from the benchmark collection to the individual time series.

\subsection{Per-instance algorithm selection and meta-learning}

This formulation is closely related to the classical algorithm-selection
problem of Rice, in which instance characteristics are mapped to the algorithm
expected to perform best on that instance \citep{rice1976}. Modern algorithm
selection systems operationalise this idea through machine learning over
instance-level meta-features, and surveys by Kotthoff and by Kerschke et al.
show that such per-instance selection can exploit complementary strengths of
algorithms that are obscured by aggregate rankings
\citep{kotthoff2014,kerschke2019}. A central requirement is therefore a feature
representation that captures properties relevant to relative algorithm
performance. Automated algorithm recommendation has also received considerable attention in time series analysis \citep{Prudncio2004, Talagala2023}. For example, \cite{Prudncio2004} employed meta-learning to identify the best-performing algorithm for TSF, whereas \cite{Shahoud2020} developed DSTMS to support algorithm selection in energy forecasting tasks.

Our recommendation target follows this paradigm directly. Each labelled
benchmark time series is an instance; the portfolio contains \cpssd{} and
\kbkssd{}; and the target label identifies which method has the smaller
absolute error relative to the reference steady-state index. Unlike conventional
algorithm-selection studies, where instance descriptors often encode a static
problem definition, here the descriptors are extracted from the observed
execution-time trajectory itself.

\subsection{Time-series representation and structure-aware recommendation}

TSFresh provides a large, automatically generated set of time-series
descriptors spanning distributional statistics, autocorrelation, entropy,
frequency-domain coefficients, peaks, quantiles, and change-related properties
\citep{PincusApproximate1991,christ2018}. This makes it suitable for converting variable-length
benchmark traces into a common fixed-dimensional representation without
hand-designing a small domain-specific feature set.

The proposed pipeline combines this direct representation with unsupervised
structure discovery. PCA is used to obtain lower-dimensional linear
representations \citep{jolliffe2002}; UMAP provides a nonlinear manifold
embedding \citep{mcinnes2018}; and HDBSCAN identifies density-based clusters
while retaining an explicit noise class \citep{campello2015}. The discovered
clusters are not treated as prediction targets. Instead, training labels are
used to characterise whether particular clusters exhibit CP-SSD or KB-KSSD
specialisation, and these specialised regions are transformed into
cluster-signature similarity features.

The final classifier is a Random Forest \citep{breiman2001}. We evaluate both a
direct TSFresh model and a hybrid model that augments selected TSFresh features
with the cluster-signature representation. This ablation is important because
it separates two questions: whether the original time-series descriptors are
sufficient for SSD recommendation, and whether the unsupervised structure adds
incremental predictive information. As shown later, the signatures expose
interpretable method-specialised structure, but the present representation does
not yield a statistically supported accuracy improvement over the direct
TSFresh classifier.

\section{Methodology}\label{sec:methodology}

This section formalises the complete analysis pipeline, from the benchmark time
series and the construction of the recommendation labels to the clustering
configuration score, cluster-signature representation, hybrid feature vector,
and final supervised recommender. All feature-selection and label-dependent
signature operations used for prediction are performed using the training
portion of each repeated holdout split.

Figure~\ref{fig:workflow} provides a visual roadmap of the methodology.
The dashed regions in the figure correspond directly to the main stages
described in this section. Section~\ref{sec:dataset} defines the benchmark
input and recommendation target; Section~\ref{sec:feature-extraction}
describes the TSFresh representation, preprocessing, and direct
predictive-feature branch; Section~\ref{sec:clustering} describes the
PCA--UMAP--HDBSCAN structure-discovery branch and the construction of
cluster signatures; and Section~\ref{sec:recommendation} combines the two
branches into the final hybrid Random Forest recommender.
Sections~\ref{sec:setup} and~\ref{sec:statistics} then define the
experimental and statistical evaluation of the resulting recommendations.

\subsection{Benchmark Dataset and Recommendation Target}\label{sec:dataset}

The available benchmark collection contains
$N_{\mathrm{all}}=5860$ cached time series. We denote series $i$ by
$\mathbf{x}_i$, its length by $L_i$, and its measurement at time index $t$ by
$x_{it}$. The labelled recommendation subset contains $N=407$ \jagt{} time
series. For each labelled series $i$, let $a_i^{\mathrm{ref}}$ denote the
reference steady-state index, while $a_i^{\mathrm{CP}}$ and
$a_i^{\mathrm{KB}}$ denote the indices returned by \cpssd{} and \kbkssd{},
respectively. An index equal to $-1$ means that the corresponding method did
not identify a steady state.

For a method $A\in\{\mathrm{CP},\mathrm{KB}\}$ that returns a valid index, its
absolute detection error is
\begin{align}
e_i^{A}=\left|a_i^{A}-a_i^{\mathrm{ref}}\right|.
\end{align}
The separate-wins formulation used in the implementation assigns binary win
indicators $w_i^{\mathrm{CP}}$ and $w_i^{\mathrm{KB}}$. If exactly one method
returns a valid index, that method receives one win. If both methods return
valid indices, the method with the smaller absolute error receives one win. If
both fail or the valid errors are equal, neither method receives a win. The
supervised target is defined only for decisive observations satisfying
$w_i^{\mathrm{CP}}+w_i^{\mathrm{KB}}=1$:
\begin{align}
y_i = w_i^{\mathrm{CP}}
\in\{0,1\},
\end{align}
where $y_i=1$ denotes a \cpssd{} win and $y_i=0$ denotes a \kbkssd{} win.
The resulting labelled dataset therefore consists of the $N=407$ pairs
$(\mathbf{x}_i,y_i)$.

In the analysed dataset, all 407 \jagt{} time series were decisive under this
formulation. \kbkssd{} wins on 255 series, while \cpssd{} wins on 152 series. The benchmark dataset composition is summarized in Table~\ref{tbl:dataset-summary}, while the outcome reasons are presented in Table~\ref{tbl:separate-wins}.

\begin{table}%[]
\caption{Summary of the benchmark dataset.}\label{tbl:dataset-summary}
\begin{tabular*}{\tblwidth}{@{}LR@{}}
\toprule
Property & Value \\
\midrule
Total number of time series & 5860 \\
Number of JAGT time series & 407 \\
CP-SSD wins & 152 \\
KB-KSSD wins & 255 \\
Non-decisive cases & 0 \\
\bottomrule
\end{tabular*}
\end{table}

\begin{table}%[]
\caption{Outcome reasons under the separate-wins formulation.}\label{tbl:separate-wins}
\begin{tabular*}{\tblwidth}{@{}LR@{}}
\toprule
Outcome reason & Count \\
\midrule
KB lower absolute error & 251 \\
CP lower absolute error & 147 \\
CP valid, KB failed & 5 \\
KB valid, CP failed & 4 \\
Both failed & 0 \\
\bottomrule
\end{tabular*}
\end{table}

\subsection{TSFresh Representation and Preprocessing}\label{sec:feature-extraction}

After removing the first observation and normalising each time series, TSFresh
maps the variable-length input to a fixed-dimensional descriptor vector
$\mathbf{z}_i=(z_{i1},\ldots,z_{id})^{\mathsf T}$, where $d$ is the number of
cached Efficient TSFresh descriptors. For any analysed collection of $n$
series, stacking these vectors yields the feature matrix
$\mathbf{Z}\in\mathbb{R}^{n\times d}$.
The descriptors include distributional statistics, entropy and complexity
measures, autocorrelation, peak and strike statistics, quantile-based change
features, and frequency-domain coefficients.

Before dimensionality reduction, each TSFresh feature is standardised using
the mean and standard deviation estimated from the data on which the
corresponding preprocessing operation is fitted. Columns that are constant in a
training split are removed before supervised feature selection. The exact
standardisation transformation is given in
Appendix~\ref{app:standardisation}.

For the supervised branch, candidate TSFresh features are ranked by their
estimated mutual information with the binary recommendation target. Within
every training split, the top $p=100$ descriptors are retained for the
best-performing model. The standard mutual-information definition used for
feature ranking is given in Appendix~\ref{app:mutual-information}. Denoting their index set by
$\mathcal{J}_p$, the selected direct TSFresh vector is
\begin{align}
\mathbf{b}_i
=
( z_{ij} )_{j\in\mathcal{J}_p}
\in\mathbb{R}^{p}.
\end{align}

% \begin{figure}[t]
% \centering
% \begin{tikzpicture}[
%     node distance=0.6cm,
%     box/.style={rectangle, rounded corners, draw, align=center, minimum width=2cm, minimum height=0.75cm, font=\small},
%     arrow/.style={-{Latex[length=2mm]}, thick}
% ]
% \node[box] (raw) {Raw benchmark\\time series};
% \node[box, right=of raw] (tsfresh) {TSFresh\\descriptors};
% \node[box, right=of tsfresh] (matrix) {Feature\\matrix};
% \node[box, below=of tsfresh] (families) {Autocorrelation\\Entropy\\Peaks\\Quantiles\\FFT\\Change statistics};

% \draw[arrow] (raw) -- (tsfresh);
% \draw[arrow] (tsfresh) -- (matrix);
% \draw[arrow] (families) -- (tsfresh);
% \end{tikzpicture}
% \caption{Overview of the TSFresh feature extraction process.}
% \label{fig:tsfresh-extraction}
% \end{figure}

% The transformation diagram summarizing the transformation from the raw
% benchmark time series to the fixed-dimensional TSFresh representation is given in
% Figure~\ref{fig:tsfresh-extraction}.

\subsection{PCA--UMAP--HDBSCAN Clustering}\label{sec:clustering}

The clustering sweep evaluates both PCA--HDBSCAN and
PCA--UMAP--HDBSCAN configurations. The primary analysis fits the dimensionality-reduction and clustering models to the 407 labelled JAGT time series so that the clustering population and recommendation population coincide. A secondary robustness analysis repeats the downstream recommendation using a clustering configuration learned from the complete collection of 5,860 cached time series.

\subsubsection{Principal component analysis}
PCA is applied to the standardised TSFresh feature matrix using the standard
covariance-eigenvector formulation \citep{jolliffe2002}. The retained principal
components provide the representation used either directly by HDBSCAN or as
input to UMAP. For the PCA-only sweep,
$p_{\mathrm{PCA}}\in\{10,20,30,50,75,100\}$, whereas configurations followed
by UMAP use $p_{\mathrm{PCA}}\in\{20,50,100\}$. The standard PCA formulation
used in the analysis is given in Appendix~\ref{app:pca-details}.

\subsubsection{UMAP embedding and HDBSCAN partition}
For the nonlinear branch, UMAP \citep{mcinnes2018} is applied to the PCA
representation with embedding dimension
$p_{\mathrm{U}}\in\{5,10,20\}$, neighbourhood size
$n_{\mathrm{nbr}}\in\{10,15,30\}$, and minimum-distance parameter
$\delta\in\{0,0.05,0.1\}$. Euclidean distance is used by UMAP. Because UMAP
employs stochastic optimisation, its random seed is fixed to 42 to make the
embeddings and clustering sweep reproducible; the seed is not treated as a
tunable hyperparameter.

HDBSCAN \citep{campello2015} is then applied either to the PCA representation
or to the UMAP embedding using Euclidean distance and minimum cluster size
$m_{\mathrm{cs}}\in\{10,20,30,50\}$. Let $c_i$ denote the HDBSCAN cluster label assigned to observation $i$.
The label $c_i=-1$ denotes an outlier, while the remaining labels identify
the $K$ non-outlier clusters. Because
HDBSCAN does not minimise a $k$-means-type sum of squared distances, each
complete clustering configuration is evaluated using the scoring function
introduced in the next section.

\subsubsection{Internal clustering quality}
Let $N_{\mathrm{clust}}$ denote the number of observations in the
population used to fit a given clustering configuration,
$\mathcal{I}=\{i:c_i\neq-1\}$ the set of non-outlier observations, and
$n_{\mathrm{out}}$ the number of outliers. We define clustering coverage as
$C_{\mathrm{cov}}=1-n_{\mathrm{out}}/N_{\mathrm{clust}}$.

Geometric clustering quality is evaluated on the non-outlier
observations using the standard Silhouette coefficient
$S_{\mathrm{sil}}$ \citep{vrezankova2018different} and the
Davies--Bouldin index (DBI) \citep{anfossi2026easy}. Their standard
definitions are given in Appendix~\ref{app:clustering-metrics}.

The implementation in this work transforms the metrics into
\begin{align}
S_{+}&=\max\{S_{\mathrm{sil}},0\},\label{eq:splus}\\
D_{+}&=\frac{1}{1+\mathrm{DBI}},\label{eq:dplus}\\
K_{+}&=\frac{\min\{K,20\}}{20},\label{eq:kplus}
\end{align}

and defines the internal clustering score as
\begin{align}
Q_{\mathrm{internal}}
=
0.45S_{+}
+0.25C_{\mathrm{cov}}
+0.20D_{+}
+0.10K_{+}.
\label{eq:internal-score}
\end{align}

By construction, $S_{+}$, $C_{\mathrm{cov}}$, and $K_{+}$ lie in
$[0,1]$, while $D_{+}\in(0,1]$. Because the coefficients in
Eq.~\ref{eq:internal-score} are non-negative and sum to one,
$Q_{\mathrm{internal}}$ is consequently bounded to $[0,1]$.

The weights in Eq.~\ref{eq:internal-score} are heuristic design weights rather
than parameters estimated from the classification results. Their
purpose is to encode the relative importance of the four properties considered
desirable for the exploratory clustering. Silhouette score receives the largest
weight 0.45, because separation from neighbouring clusters and cohesion
within a cluster constitute the primary geometric criterion. Coverage receives
weight 0.25 to prevent configurations obtaining a high geometric score by
assigning a large fraction of difficult observations to the HDBSCAN noise
class. The transformed Davies--Bouldin index contributes a further 0.20 as a
complementary measure of within-cluster compactness relative to
between-cluster separation. Finally, the number-of-clusters term receives only
0.10. It therefore acts as a weak regularising preference against degenerate
solutions with very few clusters rather than rewarding fragmentation of the
data. Capping this term at 20 clusters further prevents an arbitrarily large
number of clusters from increasing the score.

\subsubsection{Method-specialisation signal and configuration score}

For every usable non-outlier cluster containing at least five \jagt{} series,
the numbers of non-tied \kbkssd{} and \cpssd{} wins are compared with a
two-sided binomial test under equal win probability. Benjamini--Hochberg
adjustment is applied across cluster-level tests, and a cluster is considered
significant when its adjusted value is below $0.05$ and it contains at least
five non-tied wins.

Let $F_{\mathrm{best}}$ and $F_{\mathrm{orig}}$ denote the fractions of usable
\jagt{} series contained in significant clusters for, respectively, the best
and original \cpssd{} configurations. Let $n_{\mathrm{best}}$ and
$n_{\mathrm{orig}}$ be the corresponding numbers of significant clusters. For usable cluster $k$, let
$n_{k,\mathrm{cfg}}^{\mathrm{KB}}$ and
$n_{k,\mathrm{cfg}}^{\mathrm{CP}}$ denote the non-tied \kbkssd{} and
\cpssd{} win counts used to evaluate the clustering configuration. We define
the corresponding win-rate margin as
$m_k=
(n_{k,\mathrm{cfg}}^{\mathrm{KB}}-n_{k,\mathrm{cfg}}^{\mathrm{CP}})/
(n_{k,\mathrm{cfg}}^{\mathrm{KB}}+n_{k,\mathrm{cfg}}^{\mathrm{CP}})$.
The maximum absolute margin across usable clusters is
$M_{\max}=\max_k |m_k|$.

The method-specialisation signal used by the clustering sweep is
\begin{align}
Q_{\mathrm{method}}
={}&
0.45F_{\mathrm{best}}
+0.25F_{\mathrm{orig}}
+0.15\frac{\min\{n_{\mathrm{best}},5\}}{5}
\nonumber\\
&
+0.10\frac{\min\{n_{\mathrm{orig}},5\}}{5}
+0.05M_{\max}.
\label{eq:method-signal}
\end{align}
Finally, the explicit clustering-configuration quality function is
\begin{align}
Q_{\mathrm{cluster}}
=
0.45Q_{\mathrm{internal}}
+0.55Q_{\mathrm{method}}.
\label{eq:clustering-quality}
\end{align}
Equation~\eqref{eq:clustering-quality} is the metric used to evaluate the complete
PCA--UMAP--HDBSCAN configurations in the initial sweep. While its coefficients are fixed heuristic design weights, we assign slightly greater weight to the
method-specialisation component because the purpose of the clustering stage is
not only to identify geometrically coherent groups, but also to expose regions
in which the relative performance of the SSD methods differs systematically.
At the same time, the internal-quality component retains substantial weight to
discourage partitions that exhibit apparent method specialisation but poor
geometric structure. The coefficients sum to one so that the combined score
remains on the same normalized scale as its two components.

For the subsequent signature-based recommendation experiment, candidate
clusterings are additionally ranked lexicographically by their specialisation.
Let $N_{\mathrm{CP}}$ and $N_{\mathrm{KB}}$ be the numbers of \jagt{} series in
significant CP-favouring and KB-favouring clusters. We define
$B=\min\{N_{\mathrm{CP}},N_{\mathrm{KB}}\}$ as the balanced-specialisation
criterion, $P=N_{\mathrm{CP}}N_{\mathrm{KB}}$ as its product criterion, and
$T=N_{\mathrm{CP}}+N_{\mathrm{KB}}$ as total specialised coverage. The
implementation first favours configurations containing at least one significant
cluster of each type and then orders candidates by $B$, $P$, $T$, the numbers
of CP- and KB-specialised clusters, and their respective series counts, all in
descending order.

For the primary steady-JAGT-only analysis, the selected configuration is
\texttt{pca50\_umap10\_nn30\_md0.0\_mcs20}. It contains three non-outlier clusters: two KB-favouring clusters of 144 and 84 labelled series and one CP-favouring cluster of 86 labelled series. The all-timeseries robustness analysis uses \texttt{pca50\_umap5\_nn30\_md0.0\_mcs50}.

\subsubsection{Cluster signatures}\label{sec:cluster-signatures}

For cluster $k$, let $\mathcal{C}_k^{\mathrm{tr}}$ denote the set of
labelled training observations assigned to that cluster. For original TSFresh
feature $j$, let
$\overline{z}_{kj}
=|\mathcal{C}_k^{\mathrm{tr}}|^{-1}
\sum_{i\in\mathcal{C}_k^{\mathrm{tr}}}z_{ij}$
denote its mean within the training members of cluster $k$. The descriptive
cluster z-score is
\begin{align}
g_{kj}
=
\frac{\overline{z}_{kj}-\mu_j}{\sigma_j}.
\end{align}
Features are ranked by $|g_{kj}|$. For a requested signature size $q$, let
$\mathcal{F}_k^{(q)}$ contain the $q$ highest-ranked features available in the
training split. The hybrid search evaluates
$q\in\{5,8,12,20\}$, and only clusters containing at least five training
observations are eligible to form signatures.

Within a training split, let $n_k^{\mathrm{CP}}$ and $n_k^{\mathrm{KB}}$ be the
numbers of CP- and KB-winning training observations assigned to cluster $k$.
Clusters tied in these counts are discarded. Each remaining signature is
labelled CP when $n_k^{\mathrm{CP}}>n_k^{\mathrm{KB}}$ and KB when
$n_k^{\mathrm{KB}}>n_k^{\mathrm{CP}}$. Its dominance strength is
\begin{align}
\omega_k
=
\frac{|n_k^{\mathrm{CP}}-n_k^{\mathrm{KB}}|}
{n_k^{\mathrm{CP}}+n_k^{\mathrm{KB}}}.
\end{align}
The centroid of signature $k$ in its selected raw-feature subspace is
\begin{align}
\boldsymbol{\gamma}_k
=
\frac{1}{n_k}
\sum_{i\in\mathcal{C}_k^{\mathrm{tr}}}
\mathbf{z}_{i,\mathcal{F}_k^{(q)}},
\end{align}
where
$n_k=|\mathcal{C}_k^{\mathrm{tr}}|
=n_k^{\mathrm{CP}}+n_k^{\mathrm{KB}}$
and $\mathbf{z}_{i,\mathcal{F}_k^{(q)}}$ denotes observation $i$ restricted
to the signature feature set.

\subsection{Hybrid SSD Recommendation}\label{sec:recommendation}

The recommendation task is a binary classification: given a time series, the
model estimates whether \cpssd{} ($y_i=1$) or \kbkssd{} ($y_i=0$) is expected
to produce the better steady-state estimate. The evaluated strategies include
a majority baseline, a direct supervised TSFresh model, a cluster-signature
rule, and the hybrid recommender. The reported best model is the hybrid Random
Forest.

\subsubsection{Signature similarity scores}
For observation $i$ and signature $k$, the Euclidean distance in the
signature-specific feature subspace is
$d_{ik}=\|\mathbf{z}_{i,\mathcal{F}_k^{(q)}}-\boldsymbol{\gamma}_k\|_2$.
Because signatures may contain different numbers of features, this distance is
normalised by the square root of the signature dimension and converted to a
bounded similarity. The exact score implemented is 
\begin{align}
r_{ik}
=
\omega_k
\exp\!\left(
-\frac{d_{ik}}{\sqrt{|\mathcal{F}_k^{(q)}|}}
\right).
\label{eq:signature-similarity}
\end{align}
Let $\mathcal{K}_{\mathrm{CP}}$ and $\mathcal{K}_{\mathrm{KB}}$ be the sets of
CP- and KB-labelled signatures. With the configured
\texttt{weighted\_max} aggregation, the class-specific scores are
$r_i^{\mathrm{CP}}=\max_{k\in\mathcal{K}_{\mathrm{CP}}}r_{ik}$ and
$r_i^{\mathrm{KB}}=\max_{k\in\mathcal{K}_{\mathrm{KB}}}r_{ik}$. If a class
has no available signature, its score is set to zero. The IDs of the
clusters attaining these maxima are denoted by $k_i^{\mathrm{CP}}$ and
$k_i^{\mathrm{KB}}$.

\subsubsection{Signature meta-feature vector}

The implementation converts the signature results into the following
13-dimensional meta-feature vector:
\begin{align}
\mathbf{s}_i
={}&
\bigg(
 r_i^{\mathrm{CP}},
 r_i^{\mathrm{KB}},
 r_i^{\mathrm{CP}}-r_i^{\mathrm{KB}},
 |r_i^{\mathrm{CP}}-r_i^{\mathrm{KB}}|,
 \frac{r_i^{\mathrm{CP}}}{r_i^{\mathrm{KB}}+\varepsilon},
 \frac{r_i^{\mathrm{KB}}}{r_i^{\mathrm{CP}}+\varepsilon},
 \nonumber\\
&
 k_i^{\mathrm{CP}},
 k_i^{\mathrm{KB}},
 K_{\mathrm{CP}},
 K_{\mathrm{KB}},
 W_{\mathrm{CP}},
 W_{\mathrm{KB}},
 W_{\mathrm{CP}}-W_{\mathrm{KB}}
\bigg)^{\mathsf T},
\label{eq:signature-meta-vector}
\end{align}
where $\varepsilon=10^{-12}$ prevents division by zero,
$K_{\mathrm{CP}}=|\mathcal{K}_{\mathrm{CP}}|$ and
$K_{\mathrm{KB}}=|\mathcal{K}_{\mathrm{KB}}|$ are the numbers of signatures,
and $W_{\mathrm{CP}}=\sum_{k\in\mathcal{K}_{\mathrm{CP}}}\omega_k$ and
$W_{\mathrm{KB}}=\sum_{k\in\mathcal{K}_{\mathrm{KB}}}\omega_k$ are their
total dominance strengths. The two cluster identifiers are treated as numeric
meta-features, exactly as in the implementation.

\subsubsection{Hybrid feature vector and Random Forest input}

For each split, mutual-information selection is applied only to the direct
TSFresh features, producing $\mathbf{b}_i\in\mathbb{R}^{p}$. The signature
meta-features are then concatenated without further selection. The complete
input to the classifier is therefore
\begin{align}
\mathbf{h}_i
=
\begin{pmatrix}
\mathbf{b}_i\\
\mathbf{s}_i
\end{pmatrix}
\in\mathbb{R}^{p+13}.
\label{eq:hybrid-vector}
\end{align}
For the best configuration, $p=100$, so the Random Forest receives 113 columns
per observation. The signature size $q=20$ controls how many prominent TSFresh
descriptors define each cluster signature; it is distinct from the 13 derived
meta-features in Equation~\ref{eq:signature-meta-vector}.

The final supervised learner is a Random Forest \citep{breiman2001}.
We fix the number of trees to $B_{\mathrm{RF}}=300$ for all Random Forest
experiments; this value is an implementation choice and is not selected using
the evaluation data. The final class recommendation is obtained by standard majority voting across
the 300 trees. The implemented Random Forest uses unrestricted tree depth,
minimum leaf size 2, \texttt{balanced\_subsample} class weights, random seed 42, and parallel tree construction. Thus, class weighting is recomputed on the bootstrap sample used
for each tree.

\subsubsection{Recommendation workflow}

Figure~\ref{fig:workflow} summarises the complete recommendation pipeline. The
layout separates the shared preprocessing stage from two feature-construction
branches. The left branch discovers unsupervised structure and converts
training-label cluster specialisation into signature-similarity features; the
right branch retains the TSFresh descriptors selected directly for prediction.
The two representations are concatenated only after the branch-specific
processing has been completed.

\subsection{Experimental Setup and Evaluation Protocol}\label{sec:setup}
The predictive evaluation uses repeated stratified holdout validation with $R=20$ repetitions and a test fraction of $\tau=0.30$. For repetition $r$, let $\mathcal{D}_{r}^{\mathrm{tr}}$ and $\mathcal{D}_{r}^{\mathrm{te}}$ denote the training and test sets. Feature selection, training-label cluster characterisation, signature construction, and model fitting are performed using the training portion of the corresponding split. The test portion is used only for prediction and metric computation.

Performance is evaluated using accuracy, balanced accuracy, CP-SSD precision,
CP-SSD recall, and CP-SSD F1-score, with CP-SSD treated as the positive class.
The standard definitions of these metrics are given in
Appendix~\ref{app:classification-metrics}.

Repeated-holdout performance is summarised descriptively by the mean and
standard deviation of each metric across the $R=20$ test splits. Because the same benchmark series can appear in several test sets, split-level results are treated descriptively rather than as independent statistical observations. For the primary inferential analysis, repeated held-out predictions are
aggregated into one decision per unique time series. If $\overline{p}_i$ denotes
the fraction of held-out appearances in which series $i$ is predicted as
\cpssd{}, the aggregated prediction is \cpssd{} when
$\overline{p}_i\geq 0.5$ and \kbkssd{} otherwise. This yields one prediction for
each of the 407 labelled JAGT series.

\subsection{Statistical Analysis Protocol}\label{sec:statistics}

The statistical analysis distinguishes between \emph{descriptive repeated-holdout
performance} and inferential unique-series performance. The 20 repeated
70/30 holdout splits overlap, namely a benchmark series can occur in more than one test
set. Consequently, the 20 split-level accuracies are not treated as independent
experimental units and are used only to characterise the variability of the
training/evaluation procedure. Confirmatory statistical tests are instead based
on one aggregated held-out decision for each of the 407 unique JAGT series.

\paragraph{Reference strategies.}
The natural majority strategy always recommends \kbkssd{} and is therefore
correct for 255 of 407 series, corresponding to 62.65\% accuracy. Two
stochastic reference strategies are also considered. Equiprobable random
selection has expected accuracy 50.00\%, while prevalence-matched random
selection, based on the observed class proportions of 152 CP-SSD and 255
KB-KSSD winners, has expected accuracy 53.20\%.

\paragraph{Binomial inference.}
Let $X$ denote the number of correct recommendations among the $N=407$
aggregated unique-series predictions, so that the observed accuracy is
$\widehat{p}=X/N$. Uncertainty in this proportion is quantified using the exact
Clopper--Pearson 95\% confidence interval \citep{clopper1934}. Its standard
definition is given in Appendix~\ref{app:clopper-pearson}.

To test whether the recommender performs better than a random reference with
success probability $p_0$, we use an exact one-sided binomial test. The test is
applied to the equiprobable baseline $p_0=0.5$ and the prevalence-matched
baseline $p_0=0.5320$. Its standard formulation is given in
Appendix~\ref{app:binomial-test}. These tests concern stochastic reference
strategies and therefore do not exploit the observation-wise pairing available
for the deterministic always-KB-KSSD baseline.

\paragraph{Paired comparison with always-KB-KSSD.}
The hybrid recommender and the always-KB-KSSD strategy are evaluated on the
same $N$ benchmark series and their correctness indicators are therefore
paired. Let $b$ denote the number of series on which only the hybrid is correct
and $c$ the number on which only always-KB-KSSD is correct. With
$D=b+c$ discordant pairs and $m=\min\{b,c\}$, the exact two-sided McNemar
$p$-value is
\begin{align}
p_{\mathrm{McN}}
=
\min\left\{
1,\;
2\sum_{j=0}^{m}
{D\choose j}2^{-D}
\right\}.
\label{eq:mcnemar-exact}
\end{align}
This exact conditional test is used as the primary inferential comparison with
the always-KB-KSSD baseline \citep{mcnemar1947}. The magnitude and direction of
the paired effect are summarised by the matched-pairs odds ratio
\begin{align}
\mathrm{OR}_{\mathrm{matched}}=\frac{b}{c}.
\label{eq:matched-or}
\end{align}
The asymptotic McNemar statistics and the construction of the exact odds-ratio
confidence interval are given in Appendix~\ref{app:mcnemar-details}.

\paragraph{Bootstrap uncertainty.}
Bootstrap uncertainty is evaluated for accuracy, balanced accuracy,
CP precision, CP recall, CP F1-score, and the paired accuracy gain over
always-KB-KSSD using $R_{\mathrm{boot}}=50{,}000$ replications. For the
primary series-level bootstrap, resampling is stratified by the true SSD winner:
152 CP-SSD-winning and 255 KB-KSSD-winning series are sampled independently
with replacement within their respective classes, thereby preserving the
observed class composition in every replicate.

We report percentile confidence intervals and bootstrap standard deviations.
For the paired accuracy gain over always-KB-KSSD, the same resampled series are
used for both strategies in every replicate, preserving the paired structure of
the comparison \citep{efron1993}. The corresponding bootstrap definitions are
given in Appendix~\ref{app:bootstrap-details}.

\paragraph{Project-level dependence sensitivity.}
Individual benchmark series originating from the same software project may not
be statistically independent. As a sensitivity analysis, the 407 labelled
series are partitioned into $J=30$ project groups inferred from the benchmark
identifiers. Each project-bootstrap replicate samples 30 projects with
replacement; whenever a project is selected, all of its constituent series
enter the replicate, with corresponding multiplicity if the same project is
selected more than once. Metrics and paired accuracy gains are then recomputed
on the resulting replicate, and percentile intervals are obtained using the
bootstrap procedure described in Appendix~\ref{app:bootstrap-details}.
Because project membership is inferred from the benchmark naming convention
rather than from an independently curated identifier, this analysis is treated
as a dependence-sensitivity check rather than as the principal inferential
procedure.

\paragraph{Clustering-population robustness.}
Let $\widehat{y}^{(A)}_i$ and $\widehat{y}^{(B)}_i$ denote the final
unique-series recommendations obtained from the steady-JAGT-only and
all-timeseries clustering regimes, respectively. Their agreement is quantified
using Cohen's $\kappa$, which corrects the observed agreement for agreement
expected from the predictors' marginal recommendation frequencies
\citep{cohen1960}:
\begin{align}
\kappa
=
\frac{p_o-p_e}{1-p_e}.
\label{eq:kappa}
\end{align}
The definitions of $p_o$ and $p_e$ are given in
Appendix~\ref{app:kappa-details}.

Predictive correctness under the two regimes is additionally compared using
the same exact McNemar formulation as Eq.~\ref{eq:mcnemar-exact}, with $b$ and
$c$ redefined as the numbers of series on which only one of the two regimes is
correct. A non-significant result in this comparison is interpreted as absence
of evidence for a directional difference, rather than as evidence of formal
statistical equivalence.

\section{Results}\label{sec:results}
This section presents the empirical results with respect to the four research
questions introduced in Section~\ref{sec:introduction}. We first analyse the
distribution of \cpssd{} and \kbkssd{} wins and the extent to which these
preferences are reflected in the discovered cluster structure. We then evaluate
the predictive performance of the SSD recommender against random and
majority-based reference strategies, followed by an analysis of the remaining
classification errors and the contribution of the cluster-signature features.
Finally, we assess the robustness of the recommendations to the population used
for unsupervised clustering.

\begin{figure*}[bt]
  \centering
  \includegraphics[width=1.85\columnwidth]{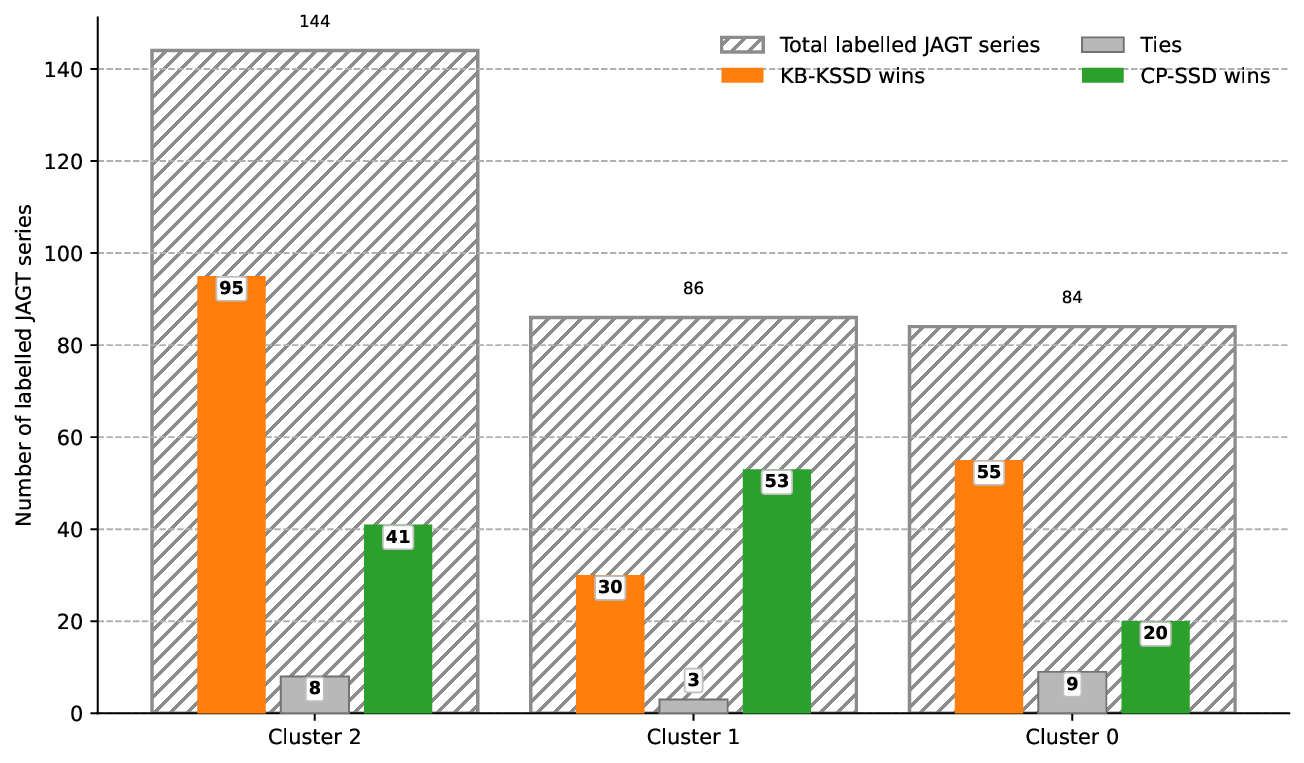}
  \caption{CP-SSD and KB-KSSD outcomes in the three non-outlier clusters of the selected steady-JAGT-only configuration. Ties in the cluster-level method comparison are shown separately and are not used to define the supervised winner label.}
  \label{fig:cluster-signatures}
\end{figure*}

\subsection{RQ1: How often does each SSD method provide the better estimate,
and does method preference vary systematically across clusters of similar time
series?}

All 407 labelled JAGT time series are decisive under the separate-wins definition. \kbkssd{} is preferred on 255 series (62.65\%), while \cpssd{} is preferred on 152 (37.35\%). Most decisions arise from smaller absolute error when both methods return valid indices: \kbkssd{} has the smaller error in 251 cases and \cpssd{} in 147. In nine additional cases, exactly one method returns a valid steady-state index, producing five \cpssd{} wins and four \kbkssd{} wins.

The selected steady-JAGT-only clustering also exposes method specialisation rather than merely reproducing the global class imbalance. Two non-outlier clusters are KB-favouring: one contains 95 \kbkssd{} wins, 41 \cpssd{} wins, and 8 ties in the cluster-level comparison, while a second contains 55 \kbkssd{} wins, 20 \cpssd{} wins, and 9 ties. The third non-outlier cluster is CP-favouring, with 53 \cpssd{} wins, 30 \kbkssd{} wins, and 3 ties. Figure~\ref{fig:cluster-signatures} visualises these distributions.

\textit{Answer to RQ1.} Both SSD methods are preferred for substantial
subsets of the labelled series, and the selected clustering contains both
KB-favouring and CP-favouring regions, demonstrating systematic
cluster-level method specialisation.

\subsection{RQ2: Can the better SSD method be recommended automatically, and
does recommendation outperform random choice and/or majority-winning detector?}

The primary hybrid recommender correctly selects the
better method for 297 of the 407 unique series, corresponding
to 72.97\% accuracy (exact 95\% CI 68.38--77.23\%).
This is 10.32 percentage points above the 62.65\% accuracy
obtained by always recommending KB-KSSD, which is correct
for 255 of 407 series (exact 95\% CI 57.75--67.37\%).
The paired comparison with always-KB-KSSD is more
informative because both strategies make a decision on the
same series. Of the 407 observations, both strategies are
correct on 228 and both are wrong on 83. The hybrid alone is
correct on 69 series, whereas always-KB-KSSD alone
is correct on only 27. The exact two-sided McNemar
test therefore rejects equality of the paired error rates
($p = 2.148 \times 10^{-5}$). The corresponding uncorrected statistic is
$\chi^2 = 18.375$ ($df = 1$, $p = 1.814 \times 10^{-5}$), and the
continuity-corrected statistic is $\chi^2 = 17.510$ ($df = 1$,
$p = 2.857 \times 10^{-5}$). The matched-pairs odds ratio is 2.56
with an exact 95\% CI of 1.62--4.15, i.e., among discordant
cases the hybrid is approximately 2.6 times as likely to be
the strategy that is correct.

\begin{figure}[pos=!htbp]
\centering
\includegraphics[width=\columnwidth]{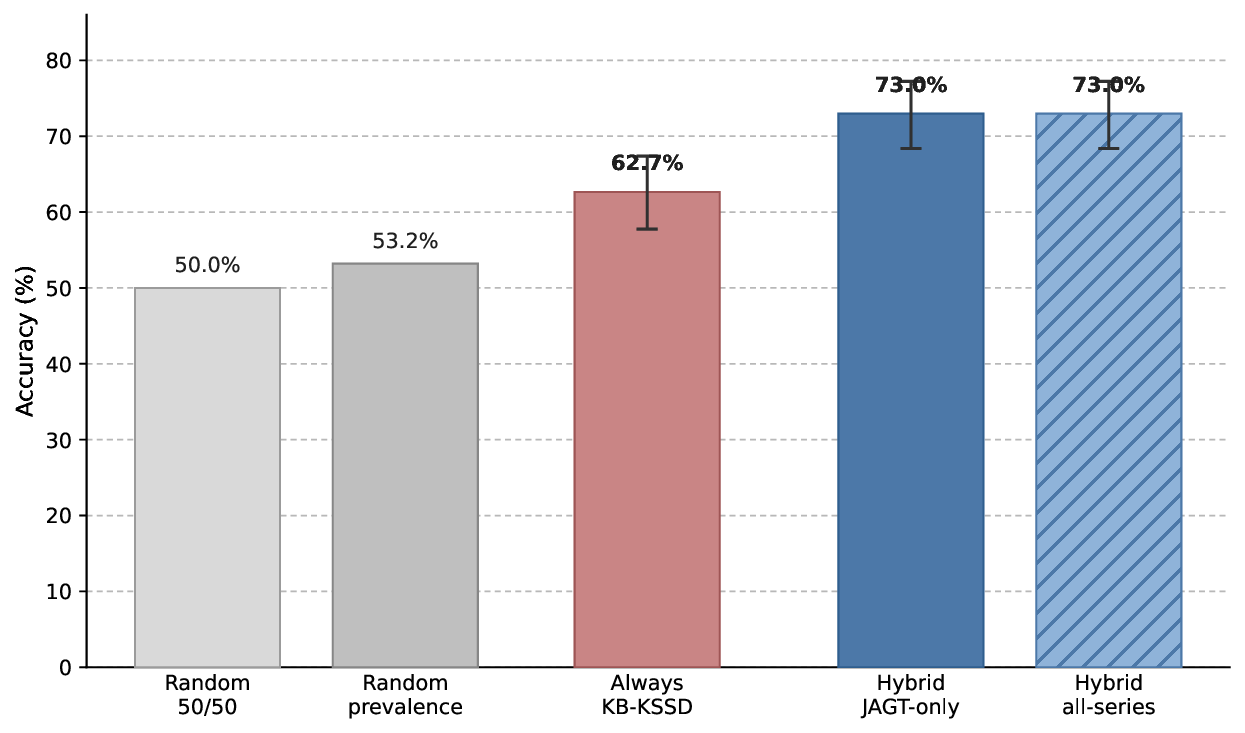}
\caption{Statistical comparison of the primary recommender with reference
strategies. The plot presents unique-series accuracy relative to random and always-KB-KSSD
baselines, with exact 95\% intervals where applicable.}
\label{fig:statistical-baselines}
\end{figure}

The recommender also exceeds both random reference strategies. Against equiprobable random selection, the exact one-sided binomial test gives $p=2.95\times10^{-21}$; against the 53.20\% prevalence-matched random expectation, $p=2.24\times10^{-16}$. These results show that the recommender is not merely exploiting
the empirical winner prevalence. Figure~\ref{fig:statistical-baselines} summarizes the unique-series accuracy of the recommender relative to the reference strategies.

\begin{figure}[pos=!htbp]
\centering
\includegraphics[width=0.48\textwidth]{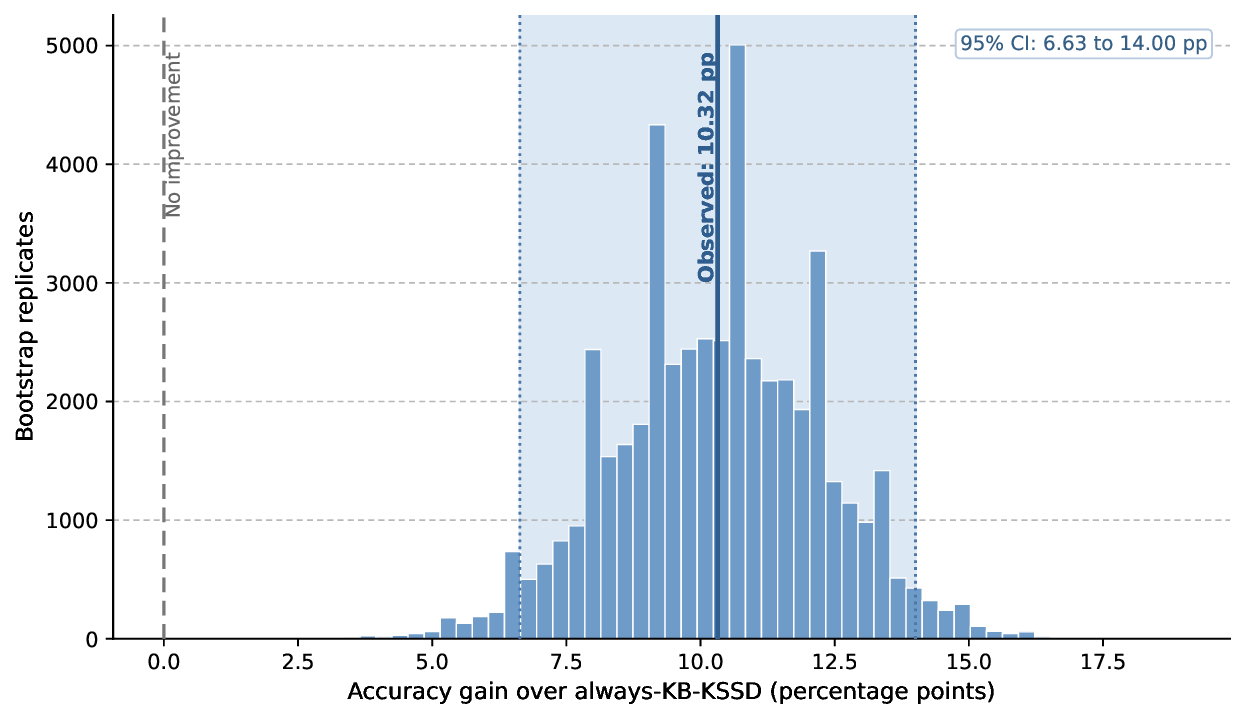}
\caption{Stratified paired-bootstrap distribution of the accuracy gain of
the primary hybrid recommender over always recommending KB-KSSD. The
distribution is based on 50,000 bootstrap replications; the observed gain is
10.32 percentage points and the 95\% percentile interval is
6.63--14.00 percentage points.}
\label{fig:bootstrap-validation1}
\end{figure}

\begin{figure}[pos=!htbp]
\centering
\includegraphics[width=0.48\textwidth]{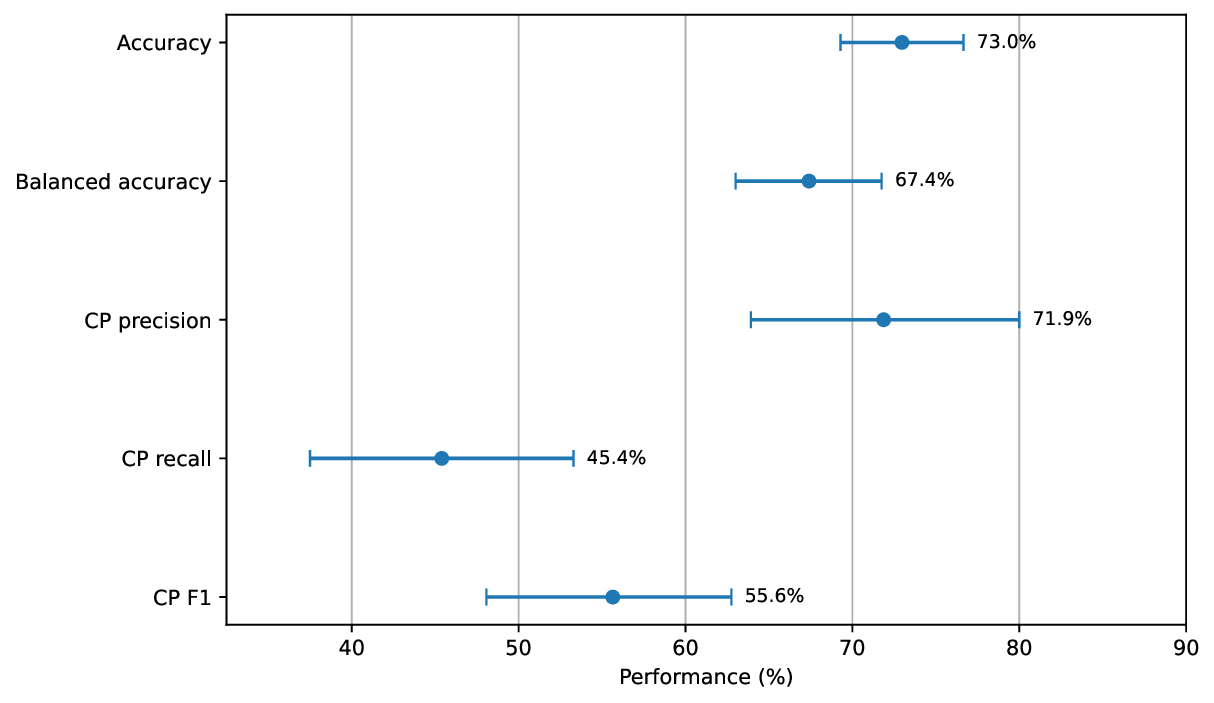}
\caption{Primary hybrid-recommender performance with 95\% percentile
intervals obtained from 50,000 class-stratified bootstrap replications.
Intervals are shown for accuracy, balanced accuracy, CP-SSD precision,
recall, and F1-score.}
\label{fig:bootstrap-validation2}
\end{figure}

The 50,000-replicate stratified paired bootstrap yields
a 95\% interval of 69.29--76.66\% for accuracy and 6.63--14.00
percentage points for the gain over always-KB-KSSD. The
corresponding bootstrap SD is 1.93 percentage points for both
accuracy and the paired accuracy gain. Thus, the entire
series-level bootstrap distribution of the paired effect remains
well above zero.
The project-cluster sensitivity analysis is wider, as expected,
but remains positive: accuracy spans 67.80--78.20\%, while
the corresponding gain interval is 5.25--15.97 percentage
points. The project-bootstrap SD is 2.66 percentage points
for accuracy and 2.74 percentage points for the paired gain.
The positive gain interval under this more dependence-aware
resampling scheme provides evidence that the improvement is
not driven only by treating series from the same project as
independent observations. The paired bootstrap distribution of the accuracy gain and the uncertainty
intervals for the principal performance metrics are visualized in
Figures~\ref{fig:bootstrap-validation1} and
\ref{fig:bootstrap-validation2}.

% \FloatBarrier % Removed for arXiv: prevents premature float flush and almost-empty pages

\begin{figure*}[t]
\centering
\includegraphics[width=0.72\textwidth]{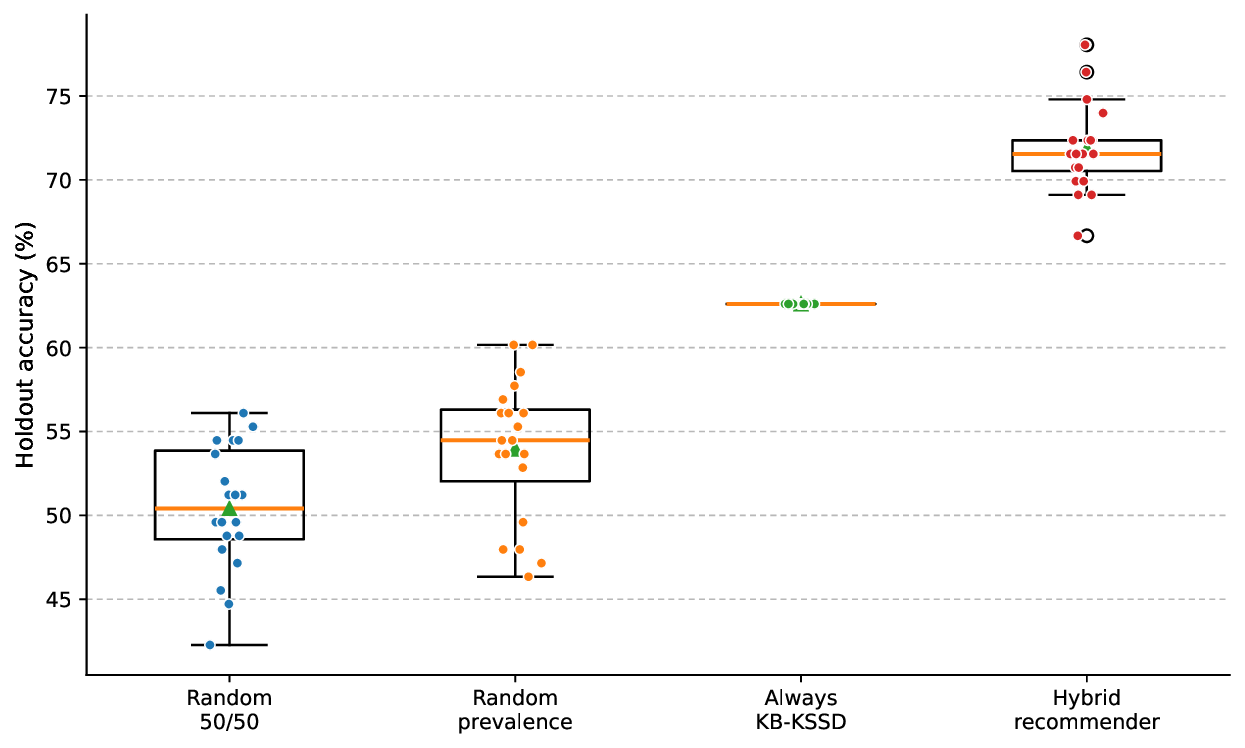}
\caption{Distribution of accuracy over the 20 repeated 70/30 holdout splits for
the primary steady-JAGT-only analysis. Points show individual split values; the
boxplot is descriptive because the test sets overlap.}
\label{fig:holdout-stability}
\end{figure*}

Across the 20 repeated 70/30 holdouts, the hybrid recommender
has mean accuracy 71.83\% (SD 2.57 percentage points),
median 71.54\%, and range 66.67--78.05\%, compared with
62.60\% for the always-KB-KSSD strategy on the same held-out
class proportions. The repeated holdouts are not used as
independent observations for hypothesis testing, but they provide
a useful stability diagnostic. Random reference strategies
fluctuate around their expected levels and remain well below the
hybrid distribution. The selected hybrid configuration and its
repeated-holdout performance are summarised in
Table~\ref{tab:best_configuration}, while Table~\ref{tab:performance} reports the corresponding
unique-series performance. The accuracy distribution is visualized in Figure~\ref{fig:holdout-stability}.

\begin{table}[t]
\centering
\caption{Configuration and repeated-holdout performance of the selected hybrid recommender. Values are means over 20 stratified 70/30 holdouts.}
\label{tab:best_configuration}
\begin{tabular}{lr}
\toprule
Property & Value \\
\midrule
Model & Random Forest \\
Feature selection & Mutual information \\
Selected TSFresh features & 100 \\
Signature size $q$ & 20 \\
Mean holdout accuracy & 71.83\% \\
Mean balanced accuracy & 66.23\% \\
Mean CP precision & 69.62\% \\
Mean CP recall & 44.02\% \\
Mean CP F1-score & 53.76\% \\
\bottomrule
\end{tabular}
\end{table}

\begin{table}[t]
\centering
\caption{Unique-series performance of the primary steady-JAGT-only hybrid recommender.}
\label{tab:performance}
\begin{tabular}{lr}
\toprule
Metric & Value \\
\midrule
Correct recommendations & 297 / 407 \\
Accuracy & 72.97\% \\
Balanced accuracy & 67.40\% \\
CP-SSD precision & 71.88\% \\
CP-SSD recall & 45.39\% \\
CP-SSD F1-score & 55.65\% \\
Predicted CP-SSD cases & 96 \\
Predicted KB-KSSD cases & 311 \\
\bottomrule
\end{tabular}
\end{table}

\subsubsection{Direct TSFresh ablation and feature-level interpretation}
To isolate the contribution of the cluster-signature meta-features, we
compare the hybrid recommender with an otherwise comparable model trained
only on the selected direct TSFresh descriptors.

The clustering-derived signatures do not produce a consistent accuracy improvement over a direct TSFresh classifier. The best direct Random Forest using mutual-information selection and 100 TSFresh features obtains 71.38\% mean holdout accuracy, compared with 71.83\% for the hybrid model, a difference of only 0.45 percentage points. When repeated held-out predictions are aggregated by the same majority rule used for the hybrid model, the direct classifier correctly predicts 299 of 407 series (73.46\%), compared with 297 (72.97\%) for the hybrid. The two unique-series predictors agree on 397 of 407 series; among the 10 disagreements, the direct model is correct in six cases and the hybrid in four (exact McNemar $p=0.754$). These results are summarized in Table~\ref{tab:ablation}. Consequently, the results do not support a claim that cluster signatures improve predictive accuracy. Their present value is instead in exposing interpretable cluster-level method specialisation and in providing a structured basis for future meta-feature designs.

\textit{Answer to RQ2.} Yes. The hybrid recommender reaches 72.97\%
unique-series accuracy and significantly outperforms both random reference
strategies and the always-KB-KSSD baseline, although it does not outperform
the direct TSFresh classifier.

\begin{table}[t]
\centering
\caption{Primary recommender and direct-TSFresh ablation. Repeated-holdout values are means over 20 splits; unique-series values aggregate repeated held-out predictions per benchmark.}
\label{tab:ablation}
\begin{tabular}{lrr}
\toprule
Metric & Direct TSFresh & Hybrid \\
\midrule
Mean holdout accuracy & 71.38\% & 71.83\% \\
Mean balanced accuracy & 65.96\% & 66.23\% \\
Unique-series accuracy & 73.46\% & 72.97\% \\
Unique balanced accuracy & 68.46\% & 67.40\% \\
Unique CP precision & 71.15\% & 71.88\% \\
Unique CP recall & 48.68\% & 45.39\% \\
Unique CP F1-score & 57.81\% & 55.65\% \\
\bottomrule
\end{tabular}
\end{table}

\begin{figure*}[t]
    \centering
    \includegraphics[width=\textwidth]{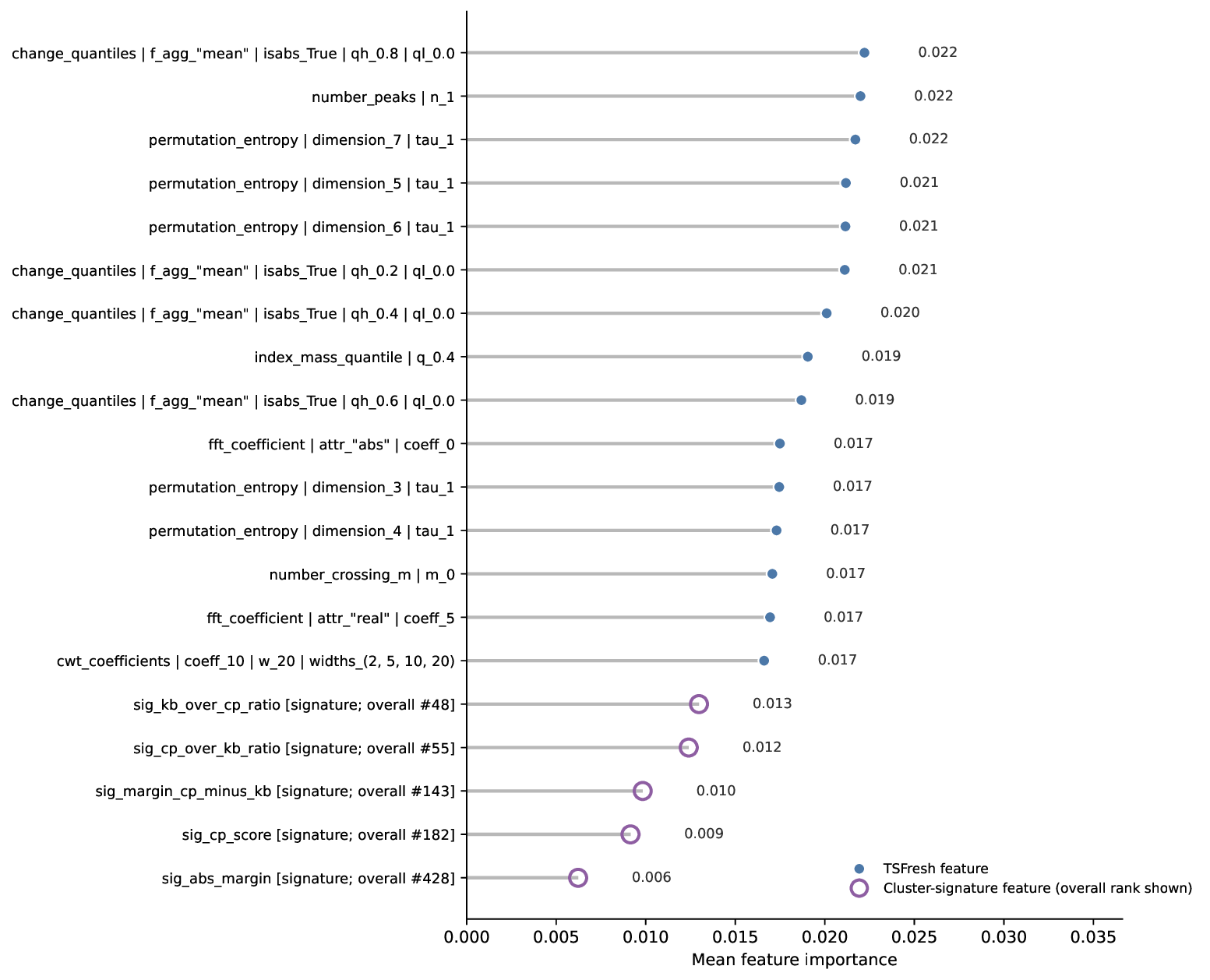}
    \caption{Highest-ranked direct TSFresh and cluster-signature features of the
primary hybrid recommendation model. For cluster-signature features, labels
also report their rank in the complete feature-importance ordering.}
    \label{fig:topFeatures}
\end{figure*}

The most influential hybrid-model variables are overwhelmingly direct TSFresh descriptors. Prominent descriptors include change-quantile statistics, peak counts, permutation entropy, index-mass quantiles, and Fourier coefficients. In the aggregated feature-importance summary, signature-derived variables account for only about 1.31\% of the reported importance mass. This is consistent with the ablation result: the cluster signatures encode meaningful structure, but the current similarity-based representation contributes little additional predictive signal beyond the direct time-series descriptors.
The most influential variables of the primary hybrid recommender are listed in Figure~\ref{fig:topFeatures}, while Table~\ref{tab:feature_importance} reports
the ten highest-ranked direct TSFresh features and their mean importances.

\begin{table*}[tbp]
\centering
\footnotesize
\caption{Ten highest-ranked TSFresh features in the aggregated feature-importance summary of the primary hybrid model.}
\label{tab:feature_importance}
\begin{tabular}{llcc}
\toprule
Rank & TSFresh feature & Mean importance & Selected in splits \\
\midrule
1 & Change quantiles ($q_l=0.0,q_h=0.8$) & 0.0222 & 1 \\
2 & Number of peaks ($n=1$) & 0.0220 & 7 \\
3 & Permutation entropy ($d=7,\tau=1$) & 0.0217 & 7 \\
4 & Permutation entropy ($d=5,\tau=1$) & 0.0212 & 14 \\
5 & Permutation entropy ($d=6,\tau=1$) & 0.0212 & 6 \\
6 & Change quantiles ($q_l=0.0,q_h=0.2$) & 0.0211 & 2 \\
7 & Change quantiles ($q_l=0.0,q_h=0.4$) & 0.0201 & 3 \\
8 & Index mass quantile ($q=0.4$) & 0.0191 & 2 \\
9 & Change quantiles ($q_l=0.0,q_h=0.6$) & 0.0187 & 7 \\
10 & FFT coefficient, absolute part ($k=0$) & 0.0175 & 7 \\
\bottomrule
\end{tabular}
\end{table*}

\subsection{RQ3: What kinds of recommendation errors remain, particularly for
the minority CP-SSD class?}
\begin{figure}[pos=!htbp]
    \centering
    \includegraphics[width=\columnwidth]{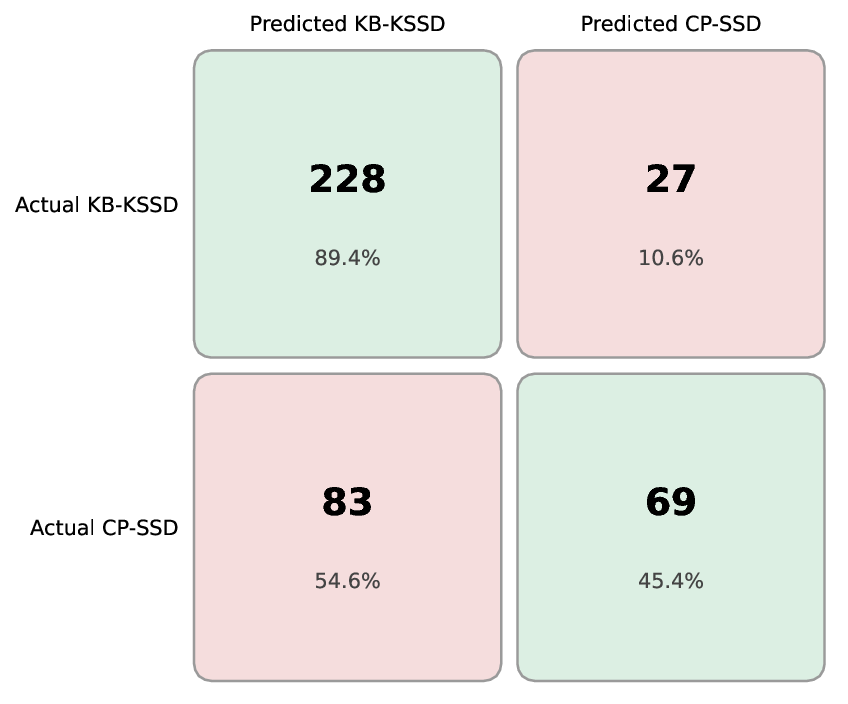}
    \caption{Confusion matrix of the steady-JAGT-only hybrid recommender.}
    \label{fig:confusionMatrix}
\end{figure}

The hybrid model is conservative in recommending the minority CP-SSD class. It predicts CP-SSD for 96 series although CP-SSD is the reference winner for 152. The unique-series confusion matrix contains 69 true positives, 27 false positives, 83 false negatives, and 228 true negatives. Thus, CP-SSD precision is 71.88\%, but recall is only 45.39\%; the corresponding F1-score is 55.65\%. In contrast, 228 of the 255 KB-KSSD-winning series are retained as KB-KSSD recommendations. Figure~\ref{fig:confusionMatrix} shows the complete confusion matrix.

The bootstrap intervals reflect the same imbalance. Under the stratified series
bootstrap, balanced accuracy has a 95\% interval of 63.01--71.75\%, CP precision
63.92--80.00\%, CP recall 37.50--53.29\%, and CP F1-score 48.07--62.75\%.
The project-cluster sensitivity analysis widens the recall interval further to
32.75--56.93\%, indicating that the principal remaining limitation is not an
inability to identify KB-KSSD cases but insufficient sensitivity to
CP-SSD-favourable executions.

\textit{Answer to RQ3.} The main remaining error mode is insufficient
sensitivity to CP-SSD-favourable series: CP-SSD recall is 45.39\%, whereas
most KB-KSSD-favourable cases are correctly retained.

\subsection{RQ4: How sensitive are the recommendations to the population used
for unsupervised clustering?}

The primary analysis learns the clustering representation only from the 407 labelled JAGT series. As a robustness check, the pipeline was also evaluated with the selected clustering configuration learned from all 5,860 cached time series. The all-timeseries variant again obtains 297 correct unique-series recommendations (72.97\%), with balanced accuracy 67.67\%, CP precision 71.00\%, and CP recall 46.71\%.

Because the overall accuracy is unchanged, the all-timeseries robustness
check also yields the same binomial results against the random references.
Its paired comparison with always-KB-KSSD contains 71 hybrid-only correct
cases and 29 always-KB-only correct cases, giving exact McNemar
$p=3.216\times10^{-5}$ and matched-pairs OR $=2.45$
(95\% CI 1.57--3.91). The complete unique-series statistical comparison for the primary and all-timeseries analyses is provided in Table~\ref{tab:exact-statistics}.

\begin{table*}[t]
\centering
\caption{Unique-series statistical validation for the primary steady-JAGT-only
regime and the all-timeseries robustness regime. Exact McNemar $p$-values are
two-sided; random-baseline binomial tests are one-sided.}
\label{tab:exact-statistics}
\begin{tabular}{lcc}
\toprule
Quantity & Steady-JAGT-only & All-timeseries \\
\midrule
Class counts (CP / KB) & 152 / 255 & 152 / 255 \\
Correct / total & 297 / 407 & 297 / 407 \\
Accuracy & 72.97\% & 72.97\% \\
Exact 95\% CI for accuracy & [68.38, 77.23]\% & [68.38, 77.23]\% \\
Balanced accuracy & 67.40\% & 67.67\% \\
CP precision & 71.88\% & 71.00\% \\
CP recall & 45.39\% & 46.71\% \\
CP F1-score & 55.65\% & 56.35\% \\
Always-KB-KSSD accuracy & 62.65\% & 62.65\% \\
Exact 95\% CI for always-KB & [57.75, 67.37]\% & [57.75, 67.37]\% \\
50/50 random expectation & 50.00\% & 50.00\% \\
Prevalence-random expectation & 53.20\% & 53.20\% \\
Gain over always-KB-KSSD & +10.32 pp & +10.32 pp \\
Both strategies correct & 228 & 226 \\
Hybrid-only correct ($b$) & 69 & 71 \\
Always-KB-only correct ($c$) & 27 & 29 \\
Both strategies wrong & 83 & 81 \\
Discordant pairs ($b+c$) & 96 & 100 \\
Exact McNemar $p$ (two-sided) & $2.148\times10^{-5}$ & $3.216\times10^{-5}$ \\
Exact McNemar $p$ (one-sided) & $1.074\times10^{-5}$ & $1.608\times10^{-5}$ \\
McNemar $\chi^2$, uncorrected & 18.375 ($p=1.814\times10^{-5}$) & 17.640 ($p=2.669\times10^{-5}$) \\
McNemar $\chi^2$, corrected & 17.510 ($p=2.857\times10^{-5}$) & 16.810 ($p=4.132\times10^{-5}$) \\
Matched-pairs OR & 2.56 [1.62, 4.15] & 2.45 [1.57, 3.91] \\
Binomial $p$ vs 50/50 random & $2.954\times10^{-21}$ & $2.954\times10^{-21}$ \\
Binomial $p$ vs prevalence random & $2.243\times10^{-16}$ & $2.243\times10^{-16}$ \\
\bottomrule
\end{tabular}
\end{table*}

More importantly, the two clustering regimes produce nearly indistinguishable
final decisions. They agree on 399 of 407 predictions (98.03\%), with Cohen's
$\kappa=0.946$. Only eight series receive different recommendations. Among
these, the steady-JAGT-only regime is correct while the all-timeseries regime is
wrong in four cases, and the reverse occurs in four cases. The exact two-sided
McNemar comparison of correctness is therefore $p=1.000$. This value arises
because the only discordant correctness outcomes are split perfectly evenly
(4 vs. 4); it does not mean that the two predictors are literally identical.
Together with the identical 72.97\% overall accuracy and the highly overlapping
bootstrap intervals, the result indicates strong empirical stability to the
clustering population, while not constituting a formal equivalence test.

The all-timeseries robustness check exhibits almost identical uncertainty.
Its stratified 95\% gain interval is +6.39--+14.25 percentage points and its
project-cluster interval is +5.19--+16.11 percentage points; the corresponding
gain SDs are 1.96 percentage points and 2.79 percentage points, respectively.
The consistency of these intervals with the primary analysis supports the
robustness of the effect to the population used for unsupervised structure
discovery.

The complete bootstrap intervals for both clustering-population analyses are presented in Table~\ref{tab:bootstrap-statistics}.

\begin{table*}[t]
\footnotesize
\centering
\caption{Bootstrap uncertainty for the hybrid recommender. Cells report the
point estimate followed by the 95\% percentile interval. ``Series'' denotes the
class-stratified paired bootstrap; ``Project'' denotes the project-cluster
sensitivity analysis.}
\label{tab:bootstrap-statistics}
\begin{tabular}{lcccc}
\toprule
Metric & JAGT Series & JAGT Project & All Series & All Project \\
\midrule
Accuracy & 72.97 [69.29, 76.66]\% & 72.97 [67.80, 78.20]\% & 72.97 [69.04, 76.90]\% & 72.97 [68.19, 77.92]\% \\
Balanced accuracy & 67.40 [63.01, 71.75]\% & 67.40 [62.15, 72.13]\% & 67.67 [63.22, 72.08]\% & 67.67 [62.72, 72.16]\% \\
CP precision & 71.88 [63.92, 80.00]\% & 71.88 [63.95, 79.22]\% & 71.00 [63.21, 78.95]\% & 71.00 [62.77, 78.95]\% \\
CP recall & 45.39 [37.50, 53.29]\% & 45.39 [32.75, 56.93]\% & 46.71 [38.82, 54.61]\% & 46.71 [34.59, 57.82]\% \\
CP F1-score & 55.65 [48.07, 62.75]\% & 55.65 [44.24, 64.15]\% & 56.35 [48.80, 63.36]\% & 56.35 [45.74, 64.34]\% \\
Gain vs always-KB & +10.32 [+6.63, +14.00] pp & +10.32 [+5.25, +15.97] pp & +10.32 [+6.39, +14.25] pp & +10.32 [+5.19, +16.11] pp \\
\bottomrule
\end{tabular}
\end{table*}

\begin{figure}[pos=!htbp]
\centering
\includegraphics[width=\columnwidth]{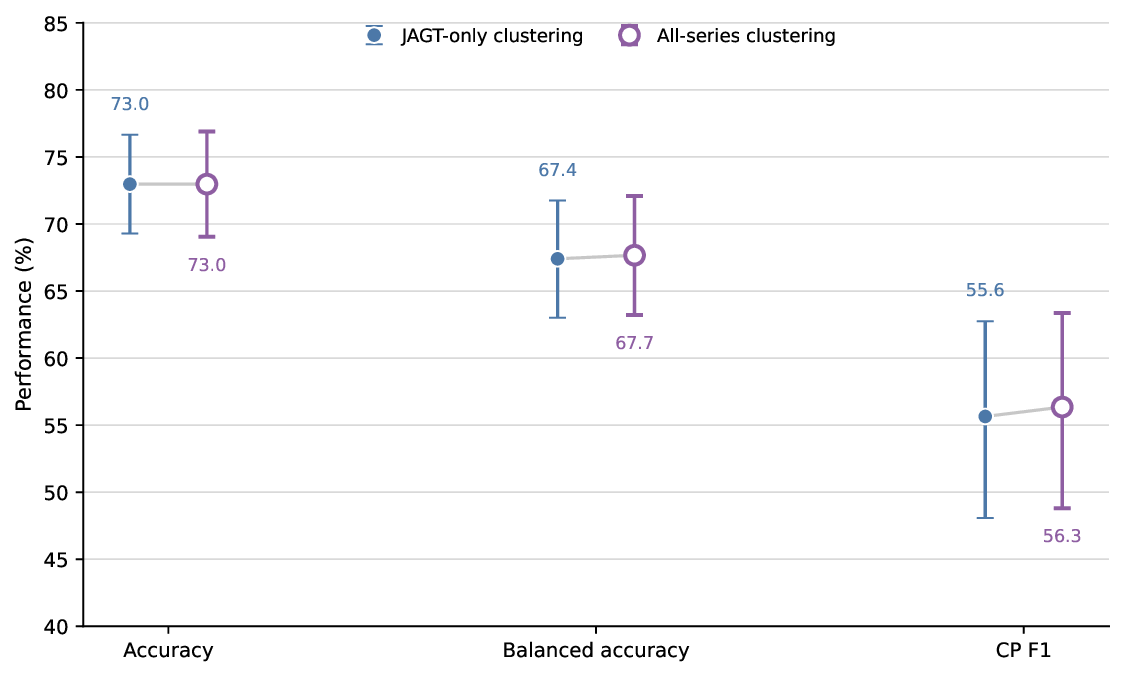}
\caption{Robustness of the final recommendation to the population used for
clustering. Both regimes reach the same unique-series accuracy and agree on
98.03\% of individual recommendations.}
\label{fig:regime-robustness}
\end{figure}

The paired comparison between the
two clustering-population regimes is summarized in Table~\ref{tab:regime-comparison}, while the comparison of the principal performance metrics is visualized in Figure~\ref{fig:regime-robustness}.

\textit{Answer to RQ4.} The recommendations are highly robust to the
clustering population: the two regimes have identical overall accuracy and
agree on 98.03\% of the individual recommendations.

\begin{table}[t]
\centering
\caption{Direct comparison of the two clustering-population regimes.}
\label{tab:regime-comparison}
\begin{tabular}{lr}
\toprule
Quantity & Value \\
\midrule
Identical predictions & 399 / 407 (98.03\%) \\
Prediction disagreements & 8 \\
Cohen's $\kappa$ & 0.946 \\
JAGT-only correct / all-series wrong & 4 \\
All-series correct / JAGT-only wrong & 4 \\
Exact McNemar $p$ & 1.000 \\
\bottomrule
\end{tabular}
\end{table}

% \FloatBarrier % Removed for arXiv: prevents premature float flush and almost-empty pages

Across the same 20 repeated 70/30 holdouts, the all-timeseries
analysis has mean accuracy 71.50\% (SD 3.49 percentage points), compared with
71.83\% (SD 2.57 percentage points) for the primary analysis. Its accuracy
ranges from 65.04\% to 77.24\%, indicating similar average performance but
somewhat greater split-to-split dispersion. The corresponding statistics for both clustering-population analyses are provided in Table~\ref{tab:holdout-stability}.

\begin{table*}[t]
\centering
\caption{Descriptive accuracy over the 20 repeated 70/30 holdout splits. The
splits overlap and therefore these values are not used as independent samples in
inferential tests.}
\label{tab:holdout-stability}
\begin{tabular}{lcccc}
\toprule
Method & JAGT mean $\pm$ SD & JAGT median [range] & All mean $\pm$ SD & All median [range] \\
\midrule
Random 50/50 & 50.41 $\pm$ 3.75 & 50.41 [42.28, 56.10] & 50.41 $\pm$ 3.75 & 50.41 [42.28, 56.10] \\
Random prevalence & 53.94 $\pm$ 4.19 & 54.47 [46.34, 60.16] & 53.94 $\pm$ 4.19 & 54.47 [46.34, 60.16] \\
Always KB-KSSD & 62.60 $\pm$ 0.00 & 62.60 [62.60, 62.60] & 62.60 $\pm$ 0.00 & 62.60 [62.60, 62.60] \\
Hybrid recommender & 71.83 $\pm$ 2.57 & 71.54 [66.67, 78.05] & 71.50 $\pm$ 3.49 & 71.14 [65.04, 77.24] \\
\bottomrule
\end{tabular}
\end{table*}

\section{Discussion}\label{sec:discussion}

The main result is that SSD selection contains learnable signal. Always choosing KB-KSSD is already a relatively strong strategy because KB-KSSD wins 62.65\% of the labelled series, yet the learned recommender raises unique-series accuracy to approximately 73\%. The paired analysis is important here: the hybrid model recovers 69 CP-SSD cases that the majority strategy necessarily misses while sacrificing only 27 KB-KSSD cases. The corresponding odds ratio and bootstrap intervals indicate that this improvement is not explained by a small number of marginal observations.

At the same time, the results refine the role of clustering in the framework. The selected clusters clearly reveal SSD specialisation, including a CP-favouring cluster despite the global KB-KSSD majority. However, the current centroid-similarity signature representation does not consistently improve prediction over direct TSFresh features. The hybrid model is slightly better on mean holdout accuracy, whereas the direct model is slightly better after unique-series aggregation, and the difference between their unique predictions is not significant. Cluster signatures should therefore be viewed as an interpretable structural layer and a starting point for richer meta-features rather than as an established source of predictive improvement.

The robustness experiment strengthens this interpretation. Learning the clustering structure from all 5,860 cached traces changes the selected PCA/UMAP/HDBSCAN configuration, yet 98.03\% of final recommendations remain unchanged. Thus, the supervised decision boundary is not strongly dependent on one particular clustering population. This stability is encouraging for replication, although it also suggests that most of the predictive signal currently comes from the direct TSFresh branch.

The principal modelling weakness is CP-SSD recall. The hybrid model identifies fewer than half of the CP-favourable series. Improving this quantity is more important than marginal gains in aggregate accuracy because an accuracy-only objective is naturally biased toward the KB-KSSD majority. Future models could therefore optimise balanced accuracy or a cost-sensitive objective, explicitly calibrate the CP decision threshold, and evaluate whether the resulting increase in CP recall provides a better practical trade-off.

\section{Threats to Validity}\label{sec:threats}
\subsection{Internal validity}

The reported model family, feature-selection strategy, signature size, and clustering configuration were selected from evaluated alternatives. Although all label-dependent feature selection and signature construction are fitted only on the training portion of each holdout, model/configuration selection and performance estimation use the same repeated-holdout experiment. The reported mean holdout performance may therefore be optimistic relative to a fully nested model-selection protocol. The unique-series aggregation reduces dependence among repeated test appearances for the final statistical comparison, but it does not remove this model-selection issue. Future work should use nested resampling or an external test collection for confirmatory evaluation.

A second threat concerns data alignment. The replication pipeline explicitly aligns extracted TSFresh rows with the canonical JAGT series keys before any label merge. This is necessary because filesystem enumeration order is not a reliable identifier. The replication package linked in Section~\ref{sec:data} therefore uses key-based alignment rather than positional matching.

\subsection{Construct validity}

The recommendation target is defined by the smaller absolute error relative to the JAGT reference steady-state index, with explicit rules for cases in which only one detector returns a valid index. This operationalisation captures location error but does not cover every property that may matter in practice, such as computational cost, stability under perturbation, or asymmetric consequences of early and late detection. Different utility functions could therefore induce different recommendation labels.

The project-cluster bootstrap uses the repository/project prefix encoded in the benchmark key and yields 30 project groups. This provides a useful sensitivity analysis for within-project dependence, but the grouping is derived from the naming convention rather than an independently curated project identifier.

\subsection{External validity}

The labelled evaluation comprises 407 JAGT Java microbenchmark time series and
focuses on two SSD methods. Although the full cache contains 5,860 series, the
additional unlabelled observations do not contribute directly to supervised
model training. The extent to which the observed results transfer to other
benchmark suites, languages, runtime environments, hardware platforms, or SSD
portfolios should therefore be investigated further. CP-SSD is also the
minority winner in the present dataset, which may contribute to the lower
recall observed for CP-SSD-favourable cases. Evaluating the recommender on
broader labelled datasets, including more CP-favourable and borderline cases,
would help clarify the generality of this behaviour.

\subsection{Conclusion validity}

The 20 repeated holdout splits overlap and are not treated as independent replicates for hypothesis testing. Inferential comparisons instead use one aggregated prediction per unique series. The exact McNemar test is appropriate for the paired comparison with always-KB-KSSD, while bootstrap intervals quantify uncertainty in effect sizes. The all-timeseries versus steady-JAGT-only comparison has only eight discordant predictions; the resulting McNemar $p=1.0$ indicates no detectable directional difference but must not be interpreted as proof of statistical equivalence.

\section{Conclusions}\label{sec:conclusions}

This paper introduced a data-driven framework for recommending steady-state detection methods for software performance benchmark time series. Using 407 decisive JAGT-labelled series, the primary hybrid recommender selects the lower-error method in 297 cases (72.97\%), improving by 10.32 percentage points over always choosing the majority KB-KSSD method. Exact paired testing and two complementary bootstrap analyses show that this improvement is statistically robust.

The experiments also identify the limits of the current design. Direct TSFresh features already contain most of the predictive signal, and the present cluster-signature representation does not yield a consistent accuracy advantage. Nevertheless, clustering exposes interpretable SSD-specialisation patterns and the downstream recommendations remain highly stable when the clustering population is expanded from the labelled JAGT subset to all 5,860 cached time series. The framework therefore supports the broader premise of automatic SSD recommendation while providing a clear direction for refinement.

Future work should: (i) expand the labelled benchmark collection, especially with CP-SSD-favourable and difficult boundary cases; (ii) evaluate cost-sensitive and threshold-calibrated recommenders; (iii) replace centroid signatures with richer distributional or soft-membership representations; (iv) exploit the unlabelled series through semi-supervised learning; (v) extend the portfolio beyond two SSD algorithms. A final confirmatory study should use nested model selection or an external benchmark collection to obtain an unbiased estimate of generalisation performance.

\section{Data and code availability}\label{sec:data}
The data and analysis materials required to reproduce the results reported in
this study are available in the accompanying versioned replication package:
\url{https://github.com/MartinBeseda/steady-state/releases/tag/v1.2.0}. The package contains the archived detector-comparison
results, cached TSFresh representation, analysis and statistical-validation
scripts, generated result tables, and figure-generation code. Upstream
CP-SSD/KB-KSSD execution scripts are also provided for provenance, although
their specialised external research dependencies are not distributed as a
self-contained software environment.

\section*{Declaration of competing interest}

The authors declare that they have no known competing financial interests or
personal relationships that could have appeared to influence the work reported
in this paper.

\section*{Declaration of generative AI and AI-assisted technologies in the manuscript preparation process}

During the preparation of this work, the authors used OpenAI ChatGPT to assist
with manuscript organisation, language and readability, and consistency
checking. After using this tool, the authors reviewed and edited the content as
needed and take full responsibility for the content of the publication.

% \section*{Acknowledgements}\label{sec:acknowledgements}

%Acknowledgements will be completed for the final submission.


\bibliographystyle{cas-model2-names}

\bibliography{cas-refs}



\appendix

\section{Supplementary Methodological Definitions}
\label{app:method-details}

\subsection{Feature Standardisation}
\label{app:standardisation}

Let $\mathcal{I}_{\mathrm{fit}}$ denote the observations on which a
preprocessing operation is fitted and let
$n_{\mathrm{fit}}=|\mathcal{I}_{\mathrm{fit}}|$. For feature $j$, define

\begin{align}
\mu_j
&=
\frac{1}{n_{\mathrm{fit}}}
\sum_{r\in\mathcal{I}_{\mathrm{fit}}} z_{rj},\\
\sigma_j
&=
\sqrt{
\frac{1}{n_{\mathrm{fit}}}
\sum_{r\in\mathcal{I}_{\mathrm{fit}}}
(z_{rj}-\mu_j)^2
}.
\end{align}

An observation $i$ is then standardised as

\begin{align}
\widetilde{z}_{ij}
=
\frac{z_{ij}-\mu_j}{\sigma_j}.
\end{align}

\subsection{Principal Component Analysis}
\label{app:pca-details}

Let $\widetilde{\mathbf{Z}}\in\mathbb{R}^{N_{\mathrm{clust}}\times d}$ be the
standardised feature matrix for the population used to fit the clustering
representation. Its empirical covariance matrix is

\begin{align}
\mathbf{C}
=
\frac{1}{N_{\mathrm{clust}}-1}
\widetilde{\mathbf{Z}}^{\mathsf T}
\widetilde{\mathbf{Z}}.
\end{align}

PCA solves

\begin{align}
\mathbf{C}\mathbf{v}_{\ell}
&=
\lambda_{\ell}\mathbf{v}_{\ell},\\
\lambda_1&\geq\lambda_2\geq\cdots\geq\lambda_d,
\end{align}

where $\lambda_{\ell}$ and $\mathbf{v}_{\ell}$ are the $\ell$-th eigenvalue
and eigenvector. For a requested dimension $p_{\mathrm{PCA}}$, the projection
matrix is

\begin{align}
\mathbf{V}_{p_{\mathrm{PCA}}}
=
(\mathbf{v}_1,\ldots,\mathbf{v}_{p_{\mathrm{PCA}}}),
\end{align}

and the PCA representation is

\begin{align}
\mathbf{P}
=
\widetilde{\mathbf{Z}}
\mathbf{V}_{p_{\mathrm{PCA}}}
\in\mathbb{R}^{N_{\mathrm{clust}}\times p_{\mathrm{PCA}}}.
\end{align}

\subsection{Silhouette and Davies--Bouldin Indices}
\label{app:clustering-metrics}

Let $\mathcal{I}=\{i:c_i\neq-1\}$ denote the set of non-outlier
observations. For $i\in\mathcal{I}$, let $a(i)$ be the mean distance from
observation $i$ to the other members of its cluster and $b(i)$ the smallest
mean distance from $i$ to another cluster. The Silhouette coefficient
\citep{vrezankova2018different} is

\begin{align}
S_{\mathrm{sil}}
&=
\frac{1}{|\mathcal{I}|}\sum_{i\in\mathcal{I}}s(i),
\end{align}

where

\begin{align}
s(i)
=
\frac{b(i)-a(i)}{\max\{a(i),b(i)\}}.
\end{align}

Let $\boldsymbol{\xi}_i$ denote the representation of observation $i$ used
for clustering, i.e., its PCA representation for PCA--HDBSCAN configurations
or its UMAP representation for PCA--UMAP--HDBSCAN configurations. For cluster
$k$, let
$\mathcal{C}_k=\{i\in\mathcal{I}:c_i=k\}$
denote its set of non-outlier members and let $\boldsymbol{\mu}_k$ be its
centroid in this representation. Its mean within-cluster dispersion is

\begin{align}
S_k
=
\frac{1}{|\mathcal{C}_k|}
\sum_{i\in\mathcal{C}_k}
\|\boldsymbol{\xi}_i-\boldsymbol{\mu}_k\|_2.
\end{align}

Let
$M_{k\ell}=\|\boldsymbol{\mu}_k-\boldsymbol{\mu}_{\ell}\|_2$
denote the distance between clusters $k$ and $\ell$. The Davies--Bouldin
index \citep{anfossi2026easy} is

\begin{align}
\mathrm{DBI}
=
\frac{1}{K}
\sum_{k=1}^{K}
\max_{\ell\neq k}
\frac{S_k+S_{\ell}}{M_{k\ell}}.
\end{align}

\subsection{Mutual Information}
\label{app:mutual-information}

For a continuous feature $Z_j$ and the binary target $Y$, mutual information is

\begin{align}
I(Z_j;Y)
=
\sum_{y\in\{0,1\}}
\int
p_{Z_j,Y}(z,y)
\log\!\left(
\frac{p_{Z_j,Y}(z,y)}
     {p_{Z_j}(z)p_Y(y)}
\right)\,\mathrm{d}z,
\end{align}

where $p_{Z_j,Y}$ is the joint density/mass function and $p_{Z_j}$ and $p_Y$
are the corresponding marginals.

\subsection{Classification Metrics}
\label{app:classification-metrics}

For a test set $\mathcal{D}^{\mathrm{te}}$ of size
$n_{\mathrm{te}}=|\mathcal{D}^{\mathrm{te}}|$, with labels $y_i$ and
predictions $\widehat{y}_i$, accuracy is
\begin{align}
\mathrm{Acc}
=
\frac{1}{n_{\mathrm{te}}}
\sum_{i\in\mathcal{D}^{\mathrm{te}}}
\mathbb{I}(y_i=\widehat{y}_i).
\end{align}

With CP-SSD treated as the positive class and
$\mathrm{TP}$, $\mathrm{TN}$, $\mathrm{FP}$, and $\mathrm{FN}$ denoting the
confusion-matrix counts,

\begin{align}
\mathrm{Precision}_{\mathrm{CP}}
&=
\frac{\mathrm{TP}}{\mathrm{TP}+\mathrm{FP}},\\
\mathrm{Recall}_{\mathrm{CP}}
&=
\frac{\mathrm{TP}}{\mathrm{TP}+\mathrm{FN}},\\
F_{1,\mathrm{CP}}
&=
\frac{
2\,\mathrm{Precision}_{\mathrm{CP}}\,
\mathrm{Recall}_{\mathrm{CP}}
}{
\mathrm{Precision}_{\mathrm{CP}}+
\mathrm{Recall}_{\mathrm{CP}}
},\\
\mathrm{BalancedAcc}
&=
\frac{1}{2}
\left(
\frac{\mathrm{TP}}{\mathrm{TP}+\mathrm{FN}}
+
\frac{\mathrm{TN}}{\mathrm{TN}+\mathrm{FP}}
\right).
\end{align}

\subsection{Clopper--Pearson Confidence Interval}
\label{app:clopper-pearson}

For $X$ successes among $N$ observations and nominal confidence level
$1-\alpha$, the exact Clopper--Pearson limits are

\begin{align}
p_{\mathrm{L}}
&=
\begin{cases}
0, & X=0,\\
B^{-1}\!\left(\frac{\alpha}{2};X,N-X+1\right), & X>0,
\end{cases}
\\
p_{\mathrm{U}}
&=
\begin{cases}
1, & X=N,\\
B^{-1}\!\left(1-\frac{\alpha}{2};X+1,N-X\right), & X<N,
\end{cases}
\end{align}
where $B^{-1}(q;a,b)$ denotes the $q$-quantile of a Beta distribution. In this
study, $\alpha=0.05$.

\subsection{Exact Binomial Test}
\label{app:binomial-test}

For a random-reference success probability $p_0$, the one-sided hypotheses are

\begin{align}
H_0:p\leq p_0,
\qquad
H_1:p>p_0.
\end{align}

If $X=x$ correct recommendations are observed among $N$ cases, the exact
one-sided binomial $p$-value is

\begin{align}
p_{\mathrm{bin}}
=
P_{H_0}(X\geq x)
=
\sum_{j=x}^{N}
{N\choose j}
p_0^j(1-p_0)^{N-j}.
\end{align}

\subsection{Additional McNemar Statistics}
\label{app:mcnemar-details}

For $D=b+c$ discordant pairs, under the null hypothesis of equal discordant
success probability,

\begin{align}
B\mid D
\sim
\mathrm{Binomial}\!\left(D,\frac{1}{2}\right).
\end{align}

For completeness, the uncorrected and continuity-corrected asymptotic McNemar
statistics are

\begin{align}
\chi^2_{\mathrm{uncorr}}
&=
\frac{(b-c)^2}{b+c},\\
\chi^2_{\mathrm{corr}}
&=
\frac{(|b-c|-1)^2}{b+c},
\end{align}

respectively, and are compared with a $\chi^2$ distribution with one degree of
freedom.

For the matched-pairs odds ratio, define

\begin{align}
\theta=\frac{b}{b+c},
\end{align}

so that

\begin{align}
\mathrm{OR}_{\mathrm{matched}}
=
\frac{\theta}{1-\theta}.
\end{align}

If $[\theta_{\mathrm{L}},\theta_{\mathrm{U}}]$ is the exact binomial confidence
interval for $\theta$, the corresponding odds-ratio interval is

\begin{align}
\left[
\frac{\theta_{\mathrm{L}}}{1-\theta_{\mathrm{L}}},
\frac{\theta_{\mathrm{U}}}{1-\theta_{\mathrm{U}}}
\right].
\end{align}

\subsection{Bootstrap Uncertainty}
\label{app:bootstrap-details}

Let

\begin{align}
T=g(\mathbf{y},\widehat{\mathbf{y}})
\end{align}

denote a scalar statistic computed from the observed labels and predictions.
For bootstrap replicate $r$ with resampled index set
$\mathcal{I}^{*(r)}$,

\begin{align}
T^{*(r)}
=
g\!\left(
\mathbf{y}_{\mathcal{I}^{*(r)}},
\widehat{\mathbf{y}}_{\mathcal{I}^{*(r)}}
\right),
\qquad
r=1,\ldots,R_{\mathrm{boot}}.
\end{align}

The percentile bootstrap confidence interval is

\begin{align}
\mathrm{CI}_{1-\alpha}^{\mathrm{boot}}
=
\left[
Q_{\alpha/2}\!\left(\{T^{*(r)}\}_{r=1}^{R_{\mathrm{boot}}}\right),
Q_{1-\alpha/2}\!\left(\{T^{*(r)}\}_{r=1}^{R_{\mathrm{boot}}}\right)
\right],
\label{eq:bootstrap-ci}
\end{align}

where $Q_q(\cdot)$ denotes the empirical $q$-quantile. The bootstrap standard
deviation is

\begin{align}
s_{\mathrm{boot}}
=
\sqrt{
\frac{1}{R_{\mathrm{boot}}-1}
\sum_{r=1}^{R_{\mathrm{boot}}}
\left(T^{*(r)}-\overline{T}^{*}\right)^2
},
\end{align}

with

\begin{align}
\overline{T}^{*}
=
\frac{1}{R_{\mathrm{boot}}}
\sum_{r=1}^{R_{\mathrm{boot}}}T^{*(r)}.
\end{align}

For the paired accuracy gain,

\begin{align}
\Delta^{*(r)}
=
\mathrm{Acc}^{*(r)}_{\mathrm{hybrid}}
-
\mathrm{Acc}^{*(r)}_{\mathrm{KB}}.
\end{align}

\subsection{Cohen's Kappa}
\label{app:kappa-details}

For predictors $A$ and $B$, the observed agreement is

\begin{align}
p_o
=
\frac{1}{N}
\sum_{i=1}^{N}
\mathbb{I}
\left(
\widehat{y}^{(A)}_i
=
\widehat{y}^{(B)}_i
\right).
\end{align}

Let

\begin{align}
q_{A,c}
&=
\frac{1}{N}
\sum_{i=1}^{N}
\mathbb{I}(\widehat{y}^{(A)}_i=c),\\
q_{B,c}
&=
\frac{1}{N}
\sum_{i=1}^{N}
\mathbb{I}(\widehat{y}^{(B)}_i=c),
\qquad c\in\{0,1\},
\end{align}

be their marginal class frequencies. The chance agreement is

\begin{align}
p_e
=
q_{A,0}q_{B,0}
+
q_{A,1}q_{B,1}.
\end{align}

\end{document}